\documentclass[preprint,10pt,a4paper]{elsarticle}

\usepackage[utf8]{inputenc}
\usepackage[T1]{fontenc}
\usepackage{amsfonts}
\usepackage{amsmath,amssymb,setspace,amsthm,fancybox,mathtools}
\usepackage{graphicx}
\usepackage{bm}
\usepackage[english]{babel}
\usepackage{geometry}
\usepackage{todonotes}
\usepackage{float}
\usepackage{epstopdf}
\usepackage{hyperref}
\usepackage{placeins}
\usepackage{booktabs}
\usepackage[modulo]{lineno}

\graphicspath{{src/figs/}} 

\theoremstyle{remark}

\newcommand{\trans}[1]{{#1}^{\mathrm{T}}}
\newcommand{\tns}[1]{\bm{\mathrm{#1}}}
\renewcommand{\vec}[1]{\bm{#1}}
\newcommand{\wei}[1]{\bm{#1}}

\newcommand{\partialder}[2]{\frac{\partial #1}{\partial #2}}
\newcommand{\norm}[1]{\left\lVert#1\right\rVert}
\newcommand{\iso}[0]{\mathrm{iso}}
\newcommand{\ani}[0]{\mathrm{ani}}

\DeclareMathOperator{\cof}{cof}

\DeclareMathOperator{\tr}{tr}

\makeatletter
\def\ps@pprintTitle{%
    \let\@oddhead\@empty
    \let\@evenhead\@empty
    \def\@oddfoot{\footnotesize\itshape
         {Submitted preprint} \hfill}%
    \let\@evenfoot\@oddfoot
    }
\makeatother

\begin{document}
	
	\begin{frontmatter}
		\title{Viscoelastic Constitutive Artificial Neural Networks (vCANNs) -- a framework for data-driven anisotropic nonlinear finite viscoelasticity}
        
        \author[1]{Kian P. Abdolazizi}
  
        \author[1]{Kevin Linka}
        
        \author[1,2]{Christian J. Cyron \corref{cor1}}
        
		\address[1]{Institute for Continuum and Material Mechanics, Hamburg University of Technology,\\ Ei\ss endorfer Stra\ss e 42, 21073 Hamburg, Germany}
        
        \address[2]{Institute of Material Systems Modeling, Helmholtz-Zentrum Hereon, \\Max-Planck-Straße 1, 21502 Geesthacht, Germany}

        \cortext[cor1]{Corresponding author: christian.cyron@tuhh.de}
		
		\begin{abstract}
            The constitutive behavior of polymeric materials is often modeled by finite linear viscoelastic (FLV) or quasi-linear viscoelastic (QLV) models. These popular models are simplifications that typically cannot accurately capture the nonlinear viscoelastic behavior of materials. For example, the success of attempts to capture strain rate-dependent behavior has been limited so far. To overcome this problem, we introduce viscoelastic Constitutive Artificial Neural Networks (vCANNs), a novel physics-informed machine learning framework for anisotropic nonlinear viscoelasticity at finite strains. vCANNs rely on the concept of generalized Maxwell models enhanced with nonlinear strain (rate)-dependent properties represented by neural networks. The flexibility of vCANNs enables them to automatically identify accurate and sparse constitutive models of a broad range of materials. To test vCANNs, we trained them on stress-strain data from Polyvinyl Butyral, the electro-active polymers VHB 4910 and 4905, and a biological tissue, the rectus abdominis muscle. Different loading conditions were considered, including relaxation tests, cyclic tension-compression tests, and blast loads. We demonstrate that vCANNs can learn to capture the behavior of all these materials accurately and computationally efficiently without human guidance.
		\end{abstract}
		
		\begin{keyword}
			Nonlinear viscoelasticity, Deep learning, Data-driven mechanics, Physics-informed machine learning, Constitutive modeling, Soft materials
        \end{keyword}
		
	\end{frontmatter}

 
	\section{Introduction}\label{sec:Intro}
    Many important materials, such as elastomers or soft biological tissues, undergo large deformations and exhibit nonlinear viscoelasticity behavior. Biological tissues additionally typically exhibit a pronounced anisotropy. Numerous experiments have confirmed that elastomers \cite{Amin2006, Haupt2001, Lion1996} and similarly ligaments and tendons exhibit nonlinear viscoelasticity \cite{Pioletti2000, Davis2012, Duenwald2009a, Miller2000, Provenzano2002, Provenzano2001, Thornton1997, Troyer2011}. In the past, many constitutive models have been proposed to characterize these materials. However, selecting an appropriate model and identifying its material parameters requires expert knowledge. Further, when selecting a model, one usually has to make a compromise between its computational efficiency and its ability to capture nonlinear viscoelasticity adequately. Therefore, this contribution aims to develop a data-driven framework that automatically discovers constitutive models for anisotropic nonlinear viscoelasticity at finite strains. Ideally, the framework is simple and numerically efficient but, at the same time, highly versatile, describing a wide range of materials. As a starting point for our development, we review existing modeling approaches and identify their advantages and limitations. 

    Broadly, existing approaches to describe viscoelasticity can be categorized into hereditary integral and internal variables models. 
    In hereditary integral models, the viscoelastic stress response is calculated by the convolution of the deformation history and an appropriate kernel function \cite{Pipkin1968,Green1957,Schapery1966}. A potential difficulty of these models is that, in particular for multiple integral models, the experimental determination of the kernel functions can be cumbersome \cite{Lockett1965, Lockett1972} and very sensitive to noise \cite{Gradowczyk1969}. Also, the numerical implementation is often challenging since one must account - in general - for the whole deformation history. Therefore, hereditary models have mainly been applied to simple one-dimensional problems and have had only a limited impact on finite element (FE) analysis. An important exception is the theory of quasi-linear viscoelasticity (QLV) \cite{Fung1981}.
    In QLV, the integrand of the hereditary integral is the product of a time-dependent reduced relaxation function and the rate of the instantaneous elastic stress, usually derived from a hyperelastic strain energy function. Often, the reduced relaxation function is represented by a Prony series \cite{Shaw2005}. Due to its computational efficiency and a large number of candidate functions for the reduced relaxation function and the instantaneous elastic stress, this approach has been used frequently \cite{Pioletti2000, Boyce2007a, Drapaca2006, Duenwald2009, Funk2000, Haut1972, Hingorani2004, Huyghe1991, Nekouzadeh2007, Puso1998, Sverdlik2002, Woo1980}. Due to the linear relationship between the reduced relaxation function and the instantaneous elastic stress within the hereditary integral, QLV falls short of representing fully general nonlinear viscoelastic behavior. In fact, normalized relaxation curves predicted by QLV have the same shape, independent of the strain, which contradicts experimental observations.
    
    Within the group of internal variable models, two model families have been particularly successful. 
    The first family follows \cite{Simo1987} and assumes an additive split of the stress into an equilibrium part and $n$ non-equilibrium overstresses, in analogy to the generalized Maxwell model. The number of overstresses is arbitrary and can be independently selected for isotropic and anisotropic contributions to the overall material behavior. Linear ordinary differential equations (ODEs) with constant coefficients govern the evolution of the overstresses, serving as internal variables. Closed-form solutions of the evolution equations by convolution integrals result in efficient time integration algorithms \cite{Holzapfel2002a}. Therefore, these models appear in many commercial FE codes \cite{Govindjee1997}. Due to the linear evolution equations, models of this family are theoretically restricted to finite linear viscoelasticity (FLV), i.e., finite strains but small perturbations away from the thermodynamic equilibrium. Although originating from different theories, FLV and QLV are similar \cite{Berjamin2021, Jridi2019}. Thus, FLV suffers from the same limitations as QLV, failing to represent general fully nonlinear viscoelastic behavior such as strain (rate)-dependent viscous properties. \cite{Pena2008} and \cite{Calvo2014} attempted to account for nonlinear viscoelastic effects by choosing strain-dependent coefficients of the evolution equation. These attempts led to improved but still not yet fully satisfactory results. FLV models have, for example, been employed in \cite{Calvo2014, Benitez2017, Gasser2011, Holzapfel2002, Kaliske1997, Pena2007a}.
    The second model family describes finite nonlinear viscoelasticity (FNLV), i.e., finite strains and finite perturbations away from the thermodynamic equilibrium \cite{Reese1998}. Motivated by the decomposition of the strain into an elastic and viscous part in the theory of linear viscoelasticity, FNLV models are based on the multiplicative decomposition of the deformation gradient into an elastic and viscous part proposed by \cite{Sidoroff1974}. In general, nonlinear ODEs govern the evolution of the viscous part of the deformation gradient, serving as an internal variable. In analogy to the generalized Maxwell model, multiple decompositions of the deformation gradient are possible, each associated with the non-equilibrium stress of a Maxwell element \cite{Reese1998} and an associated internal variable (representing a viscous part of the deformation gradient). Applications are documented for rubber \cite{Amin2006, Haupt2001, Govindjee1997, Dal2009, Scheffer2015, HooFatt2008} and soft biological tissues \cite{Nguyen2007, Liu2019a, Latorre2017, Panda2018, Bischoff2004, Latorre2015, Budday2017a}. 
    The drawbacks of FNLV models are the computational cost, especially for large-scale simulations, and their limited availability in widely used commercial FE software.

    The above models have been developed by specialists, and also the selection and calibration of these models for a specific material typically require some expert knowledge. Data-driven modeling approaches such as machine learning circumvent these problems by providing a flexible computational framework that directly infers constitutive relations from data rather than specifying them a priori \cite{Kalina2022a, Klein2022b, Hartmaier2020, Kissas2020a, Fernandez2020, Fernandez2022}.
    In deep learning, modeling the time-history effects of viscoelasticity requires a neural network for temporal signal processing. Therefore, recurrent neural networks (RNNs) and similar architectures, usually employed for speech recognition or time series prediction, have been used intensively. \cite{Oeser2009} used an Elman network to model materials with fading memory based on fractional differential equations under cyclic loading conditions. Instead of a material model, \cite{Freitag2013} applied RNNs to fuzzy data to describe time-dependent material behavior within the finite element method. The inelastic material behavior of rubber-like materials was modeled with RNNs and used in FE simulations by \cite{Zopf2017}. \cite{Chen2021} modeled small-strain viscoelasticity using long short-term memory (LSTM). RNNs compute a material's stress state based on a time window comprising the strain states of $n$ previous time steps. For FE simulations, this entails a significant increase in computational cost, as the strain states of the previous $n$ time steps have to be stored for each quadrature point. In \cite{Chen2021}, LSTMs reacted sensitively to time step sizes and loading cases, deviating from those used during training. Moreover, the width of the time window directly affects the fading memory property \cite{Truesdell2004} of the material and is difficult to determine. The thermo-viscoelastic constitutive behavior of polypropylene was modeled by \cite{Jordan2020} using a mechanistic/data-driven hybrid approach in which a neural network represented the viscous part of their rheological model.
	\cite{Marino2022} employed full-field strain measurements to calibrate the material parameters of a generalized Maxwell model for isotropic linear viscoelasticity in the small-strain regime. \cite{Linka2021a} modeled the time-dependent behavior of human brain tissue by QLV. Therein, neural networks learned the constant relaxation coefficients and times of the reduced relaxation function.
	A purely data-driven approach to constitutive modeling, which does not require any explicit constitutive models but builds on material data only, was introduced by \cite{Kirchdoerfer2016} and recently extended to inelastic materials \cite{Eggersmann2019, Salahshoor2023}. However, this approach typically requires a large database to describe the material's mechanical behavior \cite{Linka2021}. 

	To overcome the limitations of the above-delineated approaches, at least in part, herein we propose viscoelastic Constitutive Artificial Neural Networks (vCANNs), a novel physics-informed machine learning framework for anisotropic nonlinear viscoelasticity at large strains. vCANNs are based on a generalized Maxwell model, enhanced with nonlinear strain (rate)-dependent relaxation coefficients and relaxation times represented by neural networks. We show that the data-driven nature of vCANNs enables them to identify accurate anisotropic nonlinear viscoelastic constitutive models automatically. The number of Maxwell elements adapts automatically during the training, promoting a sparse model through $L_1$ regularization. Adopting the computationally very efficient framework of QLV and FLV, we leverage these well-established theories to model anisotropic nonlinear viscoelasticity. The achieved degree of accuracy is unmatched by similar traditional approaches that have been proposed. We trained vCANNs on stress-strain data from Polyvinyl Butyral, the electro-active polymers VHB 4910 and 4905, and the rectus abdominis muscle. Different loading conditions were considered, including relaxation tests, cyclic tension-compression tests, and blast loads. In all these cases, vCANNs were found to be able to learn the behavior of the materials within minutes, without human guidance, and with high accuracy.

 
    \section{Theory}\label{sec:Theory}
    In this section, we derive the theoretical framework of vCANNs. They can be considered an extension of CANNs, which were introduced in \cite{Linka2021}, to anisotropic nonlinear viscoelasticity. Therefore, we initially provide a brief review of CANNs. The section's central part will be devoted to the viscoelastic enhancement of CANNs.
    
	\subsection{Describing anisotropic hyperelasticity with generalized structural tensors}
    Many materials of interest exhibit direction-dependent, i.e., anisotropic, mechanical properties. CANNs provide a physics-informed machine learning framework for anisotropic hyperelasticity to describe these materials in a very general way. To this end, CANNs employ invariant theory, and the concept of generalized structural tensors \cite{Ehret2007}. A material is called hyperelastic if a strain energy function $\Psi=\Psi(\tns{F})$, depending only on the deformation gradient $\tns{F}$, can describe the material's mechanical behavior \cite{Basar2000}. To fulfill the principle of objectivity \cite{Ogden1985}, one usually represents $\Psi$ in terms of the right Cauchy--Green tensor $\tns{C}=\trans{\tns{F}}\tns{F}$, i.e., $\Psi=\Psi(\tns{C})$. For an incompressible hyperelastic material ($\det\tns{C} = 1$), the instantaneous elastic 2. Piola--Kirchhoff stress tensor is given by
    \begin{equation}\label{eq:2PK_incomp}
        \tns{S}^e = - p\tns{C}^{-1} + 2 \partialder{\Psi}{\tns{C}},
    \end{equation}
    where $p$ is a Lagrangian multiplier ensuring incompressibility. The superscript $(\cdot)^e$ explicitly distinguishes the instantaneous elastic stress from the viscoelastic stress derived in the next paragraph. In Eq. \eqref{eq:2PK_incomp}, the first and second terms represent the volumetric and isochoric stress contribution, respectively. The treatment of compressible and nearly compressible materials is equally possible with our proposed framework of anisotropic nonlinear viscoelasticity. For brevity, and because finite-strain viscoelasticity plays a particularly prominent role in materials often modeled as (nearly) incompressible (such as rubber materials or biological tissues), we limit the discussion in the following to incompressible materials.

    To describe the mechanical behavior of an anisotropic material, one can define several so-called preferred directions represented by unit direction vectors $\vec{l}_j \in \mathbb{R}^3$, $j=1,2,\ldots,J$, and define the following $J+1$ structural tensors
    \begin{equation}\label{eq:struc_tensor}
        \tns{L}_0=\frac{1}{3} \tns{I}, \quad \tns{L}_j = \vec{l}_j \otimes \vec{l}_j, \quad \norm{\vec{l}_j}=1, \quad j=1,2,\ldots,J.
    \end{equation}    
    Here, $\tns{I}$ denotes the second-order identity tensor, and the associated $\tns{L}_0$ is used to describe the isotropic part of the material's constitutive behavior. The preferred directions $\vec{l}_j$ can often be interpreted as directions of fiber families embedded in the material. It can be shown that to preserve the material symmetry, the strain energy function $\Psi$ has to be an isotropic function of the quantities $\tns{C}$ and $\tns{L}_j$, $j=1,2,\ldots,J$ \cite{Zhang1990}. It can be shown \cite{Itskov2009, Boehler1977} that this is the case if the strain energy function depends only on the following invariants:
    \begin{equation} \label{eq:invars_1}
        \tr\tns{C}, \quad \tr\tns{C}^2, \quad \tr\tns{C}^3, \quad \tr\left(\tns{C}\tns{L}_j\right), \quad \tr \left(\tns{C}^2\tns{L}_j\right), \quad j=1,2,\ldots,J,
    \end{equation}
    \begin{equation}\label{eq:omit_invars}
        \tr\left(\tns{C}\tns{L}_i\tns{L}_j\right), \quad \tr \left(\tns{L}_i\tns{L}_j\right),  \quad \tr \left(\tns{L}_i\tns{L}_j\tns{L}_k\right), \quad 1\le i< j < k \le J.
    \end{equation}
    The latter two types of invariants in Eq. \eqref{eq:omit_invars} are constant and can therefore be omitted from the arguments of $\Psi$. For practical applications, the influence of the first invariant type in Eq. \eqref{eq:omit_invars} is usually negligible. Therefore, $\Psi$ can commonly be expressed as
    \begin{equation}\label{eq:SEF_1}
        \Psi= \Psi\left( \tr\tns{C}, \tr\tns{C}^2, \tr\tns{C}^3, \tr\left(\tns{C}\tns{L}_1\right), \tr \left(\tns{C}^2\tns{L}_1\right), \ldots , \tr\left(\tns{C}\tns{L}_J\right), \tr \left(\tns{C}^2\tns{L}_J, \right) \right).
    \end{equation}
    With the $2R+1$ generalized invariants
    \begin{align}\label{eq:generalized_invariants_1}
        \tilde{I}_r&=\tr\left( \tns{C}\tilde{\tns{L}}_r \right) , &\tilde{J}_r&=\tr \left((\det\tns{C}) \tns{C}^{-\mathrm{T}}\tilde{\tns{L}}_r \right)=\tr \left((\cof \tns{C})\tilde{\tns{L}}_r \right) , &\mathrm{III}_{\tns{C}}&=\det\tns{C}, &r&=1,2,\ldots,R,
    \end{align}
    relying on the $R$ generalized structural tensors
    \begin{equation}\label{eq:gen_struc_tensor}
         \tilde{\tns{L}}_r= \sum_{j=0}^{J_R} w_{rj} \tns{L}_{rj}, \quad r=1,2,\ldots,R,
    \end{equation}
    where
    \begin{equation}
        \tns{L}_{r0}=\tns{L}_0, \quad \sum_{j=0}^{J_R}w_{rj}= 1, \quad w_{rj}\ge 0, \quad r=1,2,\ldots,R,        
    \end{equation}
    and employing the short-hand notation
    \begin{equation}\label{eq:invar_set_1}
        \tilde{\mathcal{I}}= \left\{ \tilde{I}_1, \tilde{J}_1, \ldots, \tilde{I}_R, \tilde{J}_R, \mathrm{III}_{\tns{C}} \right\},
    \end{equation}
    we can alternatively express Eq. \eqref{eq:SEF_1}, according to \cite{Ehret2007}, in the form
    \begin{equation}\label{eq:SEF_2}
        \Psi = \Psi \left( \tilde{\mathcal{I}} \right).
    \end{equation}
    The generalized structural tensors represent linear combinations of the standard structural tensors $\tns{L}_{rj} = \vec{l}_{rj} \otimes \vec{l}_{rj}$ introduced in Eq. \eqref{eq:struc_tensor}. We use a double index $rj$ to emphasize that,  in principle, each generalized structural tensor $\tilde{\tns{L}}_r$ can rely on a different subset of $J_r$ preferred material directions $\vec{l}_{rj}, j = 1,\ldots, J_r$.

    In order to describe not only the stress-strain behavior of a material but also the dependence of this behavior on certain in general non-mechanical parameters, it is convenient to augment the arguments of $\Psi$ with a feature vector $\tns{f} = \trans{[f_1, f_2, \cdots, f_{N_f}]}$, where $N_f$ denotes the number of features. For example, $\tns{f}$ could carry information on the material's microstructure or production process.
    Thus,
    \begin{equation}\label{eq:SEF_3}
        \Psi = \Psi \left( \tilde{\mathcal{I}}, \tns{f} \right).
    \end{equation}

    Apart from material symmetry and the principle of objectivity, the strain energy function has to fulfill several other conditions. The strain energy must always be positive, $\Psi \ge 0$. Also, the strain energy is required to approach infinity if the material is shrunk to zero or expanded to infinite volume, i.e., $\Psi \rightarrow \infty$ for $\det \tns{C} \rightarrow \infty$ or $\det \tns{C} \rightarrow 0^+$, which is called the growth condition. If a stress-free reference configuration is assumed, the strain energy function and the stress have to fulfill the normalization condition: $\Psi(\tns{C} = \tns{I}) = 0$ and $\tns{S}^e(\tns{C}=\tns{I}) = -p\tns{I} + 2\partialder{\Psi}{\tns{C}} \big|_{\tns{C}=\tns{I}} = \tns{0}$.
    Inserting Eq. \eqref{eq:SEF_3} in Eq. \eqref{eq:2PK_incomp} yields
    \begin{align}\label{eq:2_PK_general}
         \tns{S}^e &= - p\tns{C}^{-1} + \sum_{r=1}^R \underbrace{2 \left( \partialder{\Psi}{\tilde{I}_r} \tilde{\tns{L}}_r - \partialder{\Psi}{\tilde{J}_r} \tns{C}^{-1} \tilde{\tns{L}}_r \tns{C}^{-1} \right)}_{=\tns{S}_r^e} = - p\tns{C}^{-1} + \sum_{r=1}^R \; \tns{S}_r^e.
    \end{align}
    Nowadays, engineers can choose from a vast catalog of strain energy functions to model materials. Choosing a suitable strain energy function, however, typically requires expert knowledge. To overcome this problem, CANNs introduced a particularly efficient machine learning architecture to learn the relation between the argument in Eq. \eqref{eq:SEF_3} and the resulting strain energy $\Psi$. Basing CANNs on Eq. \eqref{eq:SEF_3} endows them with substantial prior knowledge from materials theory, namely, the theory of generalized invariants. This prior knowledge significantly reduces the amount of training data CANNs need to learn the constitutive behavior of a specific material of interest. At the same time, given the generality of the theory of generalized invariants, using Eq. \eqref{eq:SEF_3} as a basis does not limit the generality of CANNs in any practically relevant way. Rather the underlying neural network equips the constitutive model with the flexibility to adjust to experimental data from various materials without human guidance. In particular, the preferred material directions $\vec{l}_j$ and the scalar weight factors $w_{rj}$ in Eq. \eqref{eq:struc_tensor} and Eq. \eqref{eq:gen_struc_tensor} are learned by the CANN from the available material data. In the following, we extend this concept to anisotropic nonlinear viscoelasticity.
    
	\subsection{Viscoelasticity}
	According to \cite{Fung1981}, in QLV, the viscoelastic 2. Piola--Kirchhoff stress tensor is given by the hereditary integral
	\begin{equation}\label{eq:general_QLV}
		\tns{S}(t) = \int_{-\infty}^t \mathbb{G}(t-s) : \dot{\tns{S}}^e \mathrm{d}s,
	\end{equation}
	where $\mathbb{G}(t)$ is the time-dependent fourth-order reduced relaxation function tensor. $\dot{\tns{S}}^e$ is the material time derivative of instantaneous elastic 2. Piola--Kirchhoff stress tensor, i.e., $\dot{\tns{S}}^e = \frac{\mathrm{d}\tns{S}^e}{\mathrm{d}t}$, where $\tns{S}^e$ is computed according Eq. \eqref{eq:2_PK_general}.
 
    The fundamental assumption of QLV is that $\mathbb{G}(t)$ depends only on time but not the deformation (time-deformation separability) such that the relaxation behavior is for any applied deformation the same. To overcome this limitation, we allow $\mathbb{G}$ to depend on the deformation $\tns{C}$. At this point, we go beyond the framework of classical QLV because $\mathbb{G}$ is not only time-dependent anymore. 
    We add the deformation rate $\dot{\tns{C}}$ to the arguments of $\mathbb{G}$ since many materials show not only strain-dependent but also strain rate-dependent viscoelastic behavior \cite{Haupt2001, Hossain2020a}:
	\begin{equation}
		\tns{S}(t) = \int_{-\infty}^t \mathbb{G}(t-s; \tns{C}, \dot{\tns{C}}) : \dot{\tns{S}}^e \mathrm{d}s.
	\end{equation}
    In the simplest case, $\mathbb{G}(t)=G(t)\mathbb{I}$ where $G(t)$ denotes a scalar reduced relaxation function and $\mathbb{I}$ the fourth-order identity tensor. However, anisotropic materials may exhibit different viscous properties in different directions. Therefore, a single scalar reduced relaxation function would, in general, be insufficient to capture the complex nature of anisotropic viscoelastic materials. On the other hand, the experimental identification of a fourth-order reduced relaxation function tensor is highly challenging, even for simple classes of anisotropy, and is practically often unfeasible for complex classes. A reasonable compromise between practicability and generality of the constitutive model is to use a scalar-valued reduced relaxation function $G_r$ for each stress contribution $\tns{S}_r^e$ in Eq. \eqref{eq:2_PK_general}. Additionally, we augment the arguments of the reduced relaxation with the structural tensors to account for anisotropy:
    \begin{equation}\label{eq:reduced_stress_integral_1}
        \tns{S}(t) = - p\tns{C}^{-1} + \sum_{r=1}^R \int_{-\infty}^t G_{r}(t-s; \tns{C}, \dot{\tns{C}}, \tns{L}_1, \tns{L}_2,\ldots, \tns{L}_J) \; \dot{\tns{S}}_r^e \; \mathrm{d}s.
    \end{equation}
    Experiments suggest that in many rubber materials and soft biological tissues, the viscous effects mostly attribute to the isochoric part of the stress \cite{Simo1998}. In the incompressible limit, this holds exactly \cite{Drapaca2007}. Therefore, in Eq. \eqref{eq:reduced_stress_integral_1}, the reduced relaxation functions $G_r$ affect only the isochoric part of the stress.\\

    The reduced relaxation functions $G_r$ are scalar-valued functions of tensors. To fulfill the principle of material objectivity and to reflect the material symmetry correctly, the reduced relaxation functions have to be isotropic functions of the tensor system $\{\tns{C}, \dot{\tns{C}}, \tns{L}_1, \tns{L}_2,\ldots, \tns{L}_J \}$  \cite{Zhang1990}. Compared to Eqs. \eqref{eq:invars_1} and \eqref{eq:omit_invars}, the set of isotropic invariants, in terms of which all other isotropic functions can be expressed, is completed by \cite{Itskov2009, Boehler1977},
	\begin{equation}\label{eq:invarns_g_0}
	    \tr\dot{\tns{C}}, \quad \tr\dot{\tns{C}}^2, \quad \tr\dot{\tns{C}}^3, \quad \tr\left(\dot{\tns{C}}\tns{L}_j\right), \quad \tr \left(\dot{\tns{C}}^2\tns{L}_j\right), \quad j=1,2,\ldots,J,
	\end{equation}
    \begin{equation}\label{eq:invarns_g_2}
        \tr\left(\dot{\tns{C}}\tns{L}_i\tns{L}_j\right), \quad \tr \left(\tns{L}_i\tns{L}_j\right),  \quad \tr \left(\tns{L}_i\tns{L}_j\tns{L}_k\right), \quad 1\le i< j < k \le J,
    \end{equation}
    \begin{equation}\label{eq:invarns_g_3}
        \tr \left(\tns{C} \dot{\tns{C}} \right), \quad \tr \left( \tns{C}^2 \dot{\tns{C}} \right), \quad \tr \left( \tns{C} \dot{\tns{C}}^2 \right), \quad \tr \left( \tns{C}^2 \dot{\tns{C}}^2 \right), \quad \tr \left( \tns{C} \dot{\tns{C}} \tns{L}_j \right), \quad j=1,2,\ldots,J.
    \end{equation}
    Following the same arguments as before, we omit the invariants in Eqs. \eqref{eq:invarns_g_2} and \eqref{eq:invarns_g_3} yielding
    \begin{multline}\label{eq:reduced_relax_1}
        G_{r}= G_r \Big( t; \tr\tns{C}, \tr\tns{C}^2, \tr\tns{C}^3, \tr\left(\tns{C}\tns{L}_1\right), \tr \left(\tns{C}^2\tns{L}_1\right), \ldots , \tr\left(\tns{C}\tns{L}_J\right), \tr \left(\tns{C}^2\tns{L}_J, \right),  \\ \tr\dot{\tns{C}}, \tr\dot{\tns{C}}^2, \tr\dot{\tns{C}}^3, \ldots, \tr\left(\dot{\tns{C}}\tns{L}_1\right), \tr \left(\dot{\tns{C}}^2\tns{L}_1\right), \tr\left(\dot{\tns{C}}\tns{L}_J\right), \tr \left(\dot{\tns{C}}^2\tns{L}_J\right) \Big).
    \end{multline}
    By introducing the $2R+1$ generalized invariants
    \begin{align}\label{eq:generalized_invariants_2}
        \tilde{\dot{I}}_r&=\tr\left( \dot{\tns{C}}\tilde{\tns{L}}_r \right) , &\tilde{\dot{J}}_r&=\tr \left((\det\dot{\tns{C}}) \dot{\tns{C}}^{-\mathrm{T}}\tilde{\tns{L}}_r \right)=\tr \left((\cof\dot{\tns{C}}) \tilde{\tns{L}}_r \right) , &\mathrm{III}_{\dot{\tns{C}}}&=\det\dot{\tns{C}}, &r&=1,2,\ldots,R
    \end{align} and the short-hand notations
    \begin{align}\label{eq:invar_set_2}
        \tilde{\dot{\mathcal{I}}} &= \left\{\tilde{\dot{I}}_1, \tilde{\dot{J}}_1, \ldots, \tilde{\dot{I}}_R, \tilde{\dot{J}}_R, \mathrm{III}_{\dot{\tns{C}}} \right\}, &\mathcal{I} &= \tilde{\mathcal{I}} \cup \tilde{\dot{\mathcal{I}}},
    \end{align}
    we can express Eq. \eqref{eq:reduced_relax_1} alternatively by
    \begin{equation}\label{eq:reduced_relax_2}
        G_{r}= G_r \left( t; \mathcal{I} \right).
    \end{equation}
    Finally, we augment the arguments of the reduced relaxation function with the above-introduced feature vector $\tns{f}$:
    \begin{equation}\label{eq:reduced_relax_3}
        G_{r}= G_r \left( t; \mathcal{I}, \tns{f} \right).
    \end{equation}
    From Eq. \eqref{eq:reduced_stress_integral_1}, we obtain the 2. Piola--Kirchhoff stress tensor
    \begin{equation}\label{eq:reduced_stress_integral_2}
        \tns{S}(t) = - p\tns{C}^{-1} + \sum_{r=1}^R \int_{-\infty}^t G_r ( t-s; \mathcal{I}, \tns{f}) \; \dot{\tns{S}}_r^e \; \mathrm{d}s.
    \end{equation}
    
\begin{figure}[tbp]
		\centering
		\includegraphics[width=0.5\linewidth]{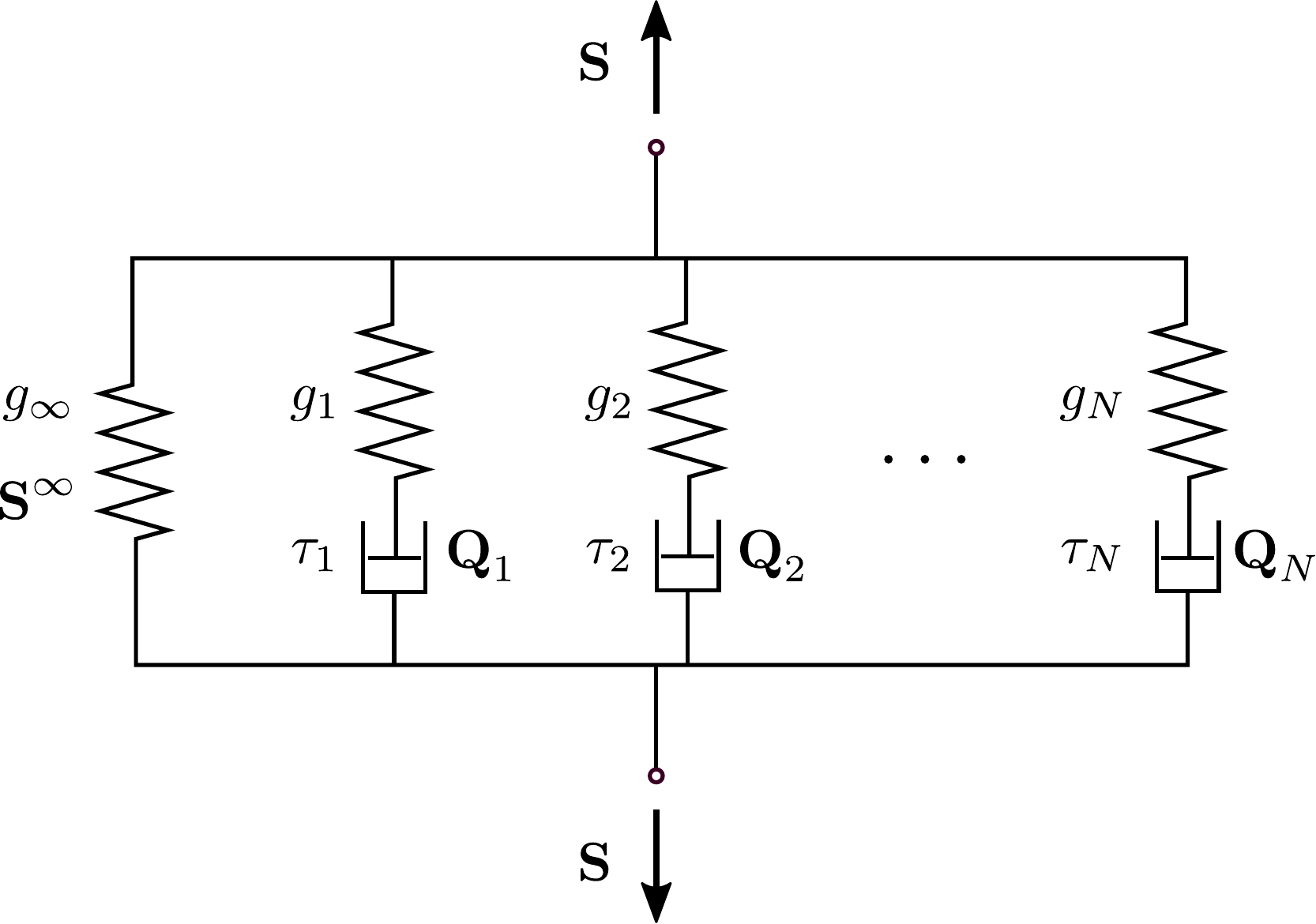}
		\caption[Generalized Maxwell model]{The generalized Maxwell model: the elastic spring on the left represents the equilibrium stress response $\tns{S}^\infty$; each Maxwell element produces a viscous overstress $\tns{Q}_\alpha$ and represents a relaxation process with a different relaxation time. $g_\infty$, $g_i$, and $\tau_i$ are constant material parameters or deformation (rate)-dependent functions.}
		\label{fig:generalizedmaxwell}
	\end{figure}

 	\paragraph{Prony Series}
    Motivated by linear viscoelasticity and the generalized Maxwell model (Fig. \ref{fig:generalizedmaxwell}), the most popular choice for the reduced relaxation function $G$ in QLV is the discrete Prony series
	\begin{align}\label{eq:Prony}
		G(t) &= g_\infty + \sum_{\alpha=1}^{N} g_\alpha \exp \left( {-\frac{t}{\tau_\alpha}} \right) 
	\end{align}
	with 
	\begin{align}\label{eq:Prony_restrictions}
		g_\infty + \sum_{\alpha=1}^{N} g_\alpha & = 1, &0\le g_\infty,g_\alpha\le 1,&&\tau_\alpha > 0.
	\end{align}
	Here, $g_\infty$ is a material parameter related to the equilibrium elasticity of the generalized Maxwell model, and $g_\alpha$ and $\tau_\alpha$ are parameters characterizing elasticity and viscous relaxation time of the $\alpha$-th Maxwell element. The $g_{\infty}$ and $g_{\alpha}$ are referred to as relaxation coefficients. In principle, the number of Maxwell elements $N$ is arbitrary, which enables the model to describe complex viscoelastic materials. Since the material parameters $g_\infty$,  $g_\alpha$, and $\tau_\alpha$ are constants, the classical Prony series is limited to linear viscoelasticity. 
	
	\paragraph{Generalized Prony Series}
    The classical Prony series does not depend on the deformation or the deformation rate but on time only. Therefore, to account for nonlinear viscoelasticity, we propose the following generalized Prony series
	\begin{equation}\label{eq:generalized_Prony}
		G_r = G_r(t;\mathcal{I}, \tns{f})= g_{r\infty}(\mathcal{I}, \tns{f}) + \sum_{\alpha=1}^{N_r} g_{r\alpha}(\mathcal{I}, \tns{f}) \exp \left( {-\frac{t}{\tau_{r\alpha}(\mathcal{I},\tns{f})}} \right).
	\end{equation}
    The conditions Eq. \eqref{eq:Prony_restrictions} individually apply to the relaxation coefficients $g_{r\infty}(\mathcal{I},\tns{f})$, $g_{r\alpha}(\mathcal{I},\tns{f})$ and the relaxation times $\tau_{r\alpha}(\mathcal{I},\tns{f})$ associated with the instantaneous elastic stress component $\tns{S}_r^e$. $N_r$ denotes the number of Maxwell branches of the generalized Maxwell model associated with the instantaneous elastic stress component $\tns{S}_r^e$. In contrast to the classical Prony series, the relaxation coefficients and times in Eq. \eqref{eq:Prony} are functions of the invariants $\mathcal{I}$ and the feature vector $\tns{f}$ to capture also anisotropic nonlinear viscoelasticity.
    
	Inserting Eq. \eqref{eq:generalized_Prony} into Eq. \eqref{eq:reduced_stress_integral_2} yields
		\begin{align}\label{eq:integral_1}
			\tns{S}(t) &=  -p\tns{C}^{-1}	
            + \sum_{r=1}^R \Bigg[ \tns{S}_r^{\infty} +  \sum_{\alpha=1}^{N_r} \underbrace{\int_{-\infty}^t g_{r\alpha}(\mathcal{I}, \tns{f}) \exp \left( {-\frac{t-s}{\tau_{r\alpha}(\mathcal{I}, \tns{f})}} \right) \dot{\tns{S}}_r^e \;\mathrm{d}s}_{=\tns{Q}_{r\alpha}} \Bigg]\\
            &=  -p\tns{C}^{-1} + \sum_{r=1}^R \Bigg[ \tns{S}_r^{\infty} +  \sum_{\alpha=1}^{N_r} \tns{Q}_{r\alpha} \Bigg]
		\end{align}
	where $\tns{S}_r^\infty = g_{r\infty}(\mathcal{I}, \tns{f}) \, \tns{S}_r^e$
	denotes the equilibrium stress associated with the $r$-th generalized Maxwell model. $\tns{Q}_{r\alpha}$ is the viscous overstress in the $\alpha$-th Maxwell branch of the $r$-th generalized Maxwell model. To illustrate the proposed constitutive model, we particularized a vCANN for the important case of transverse isotropy in \ref{sec:transverse_isotropy}. 
	In general, closed-form solutions do not exist for the integrals in Eq. \eqref{eq:integral_1} so that a numerical time integration scheme has to be applied. Details are provided 
  in \ref{sec:numericalintegration}.\\

    
	\section{Machine learning architecture} \label{sec:architecture}

	\subsection{General} 
 
\begin{figure}[]
		\centering
		\includegraphics[width=\linewidth]{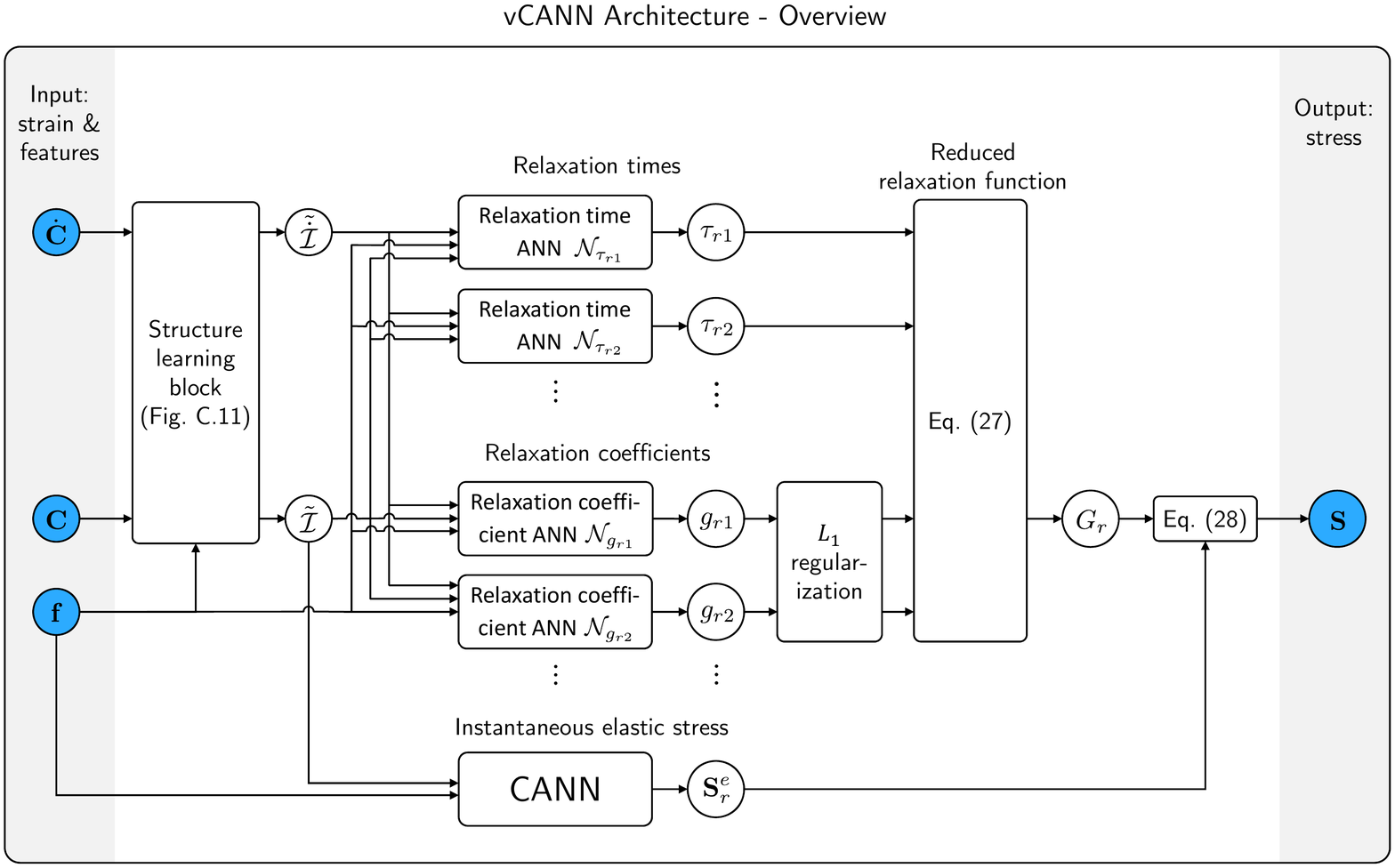}
		\caption{Schematic illustration of the vCANN architecture: The strain (rate) tensors $\tns{C}$, $\dot{\tns{C}}$, and the feature vector $\tns{f}$ serve as input to the structure learning block (Fig. \ref{fig:struc_learn}) which learns the generalized invariants $\tilde{\mathcal{I}}$ and $\tilde{\dot{\mathcal{I}}}$. The generalized invariants and the feature vector are fed to the relaxation time ANNs $\mathcal{N}_{\tau_{r\alpha}}$ and relaxation coefficient ANNs $\mathcal{N}_{g_{r\alpha}}$. The outputs of $\mathcal{N}_{\tau_{r\alpha}}$ are the relaxation times $\tau_{r\alpha}$. The neural networks $\mathcal{N}_{g_{r\alpha}}$ calculate the relaxation coefficients $g_{r\alpha}$ which are regularized to promote a sparse model (grey box `$L_1$ regularization'). The reduced relaxation function $G_r$ is obtained by inserting $\tau_{r\alpha}$ and $g_{r\alpha}$ in Eq. \eqref{eq:generalized_Prony}. The generalized invariants $\tilde{\mathcal{I}}$ and $\tns{f}$ are fed to the CANN which calculates the instantaneous elastic stress contributions $\tns{S}_r^e$. The internal structure of the CANN is depicted in Fig. 1(a) in \cite{Linka2021}. With Eq. \eqref{eq:integral_1}, we finally calculate the viscoelastic stress $\tns{S}$. Note that Fig. \ref{fig:struc_learn} is slightly modified compared to Fig. 1(b) in \cite{Linka2021} since in vCANNs one has to account for $\dot{\tns{C}}$ as an input, too.}
		\label{fig:architecture}
	\end{figure}

	In the previous section, we outlined the theoretical foundations of the model of nonlinear viscoelasticity on which we rely in this paper. The main idea of vCANNs is to implement this theory via a machine learning architecture. This architecture is illustrated in Fig. \ref{fig:architecture}.
	In our approach, we use feedforward neural networks (FFNNs). The networks consist of $H+1$ layers, that is $H$ hidden layers, of neurons. The input passed to the first layer is a vector $\wei{x}_0\in\mathbb{R}^{n_0}$. The output of the $l$-th layer is denoted by $\wei{x}_l\in\mathbb{R}^{n_l}$, respectively, and computed as
	\begin{equation}
		\wei{x}_l = \sigma_l\left( \wei{W}^{(l)} \wei{x}_{l-1} + \wei{b}^{(l)} \right), \quad l=1,\ldots,H+1, \quad \wei{x}_l\in\mathbb{R}^{n_l},
	\end{equation}
	with the activation function $\sigma_l(\cdot)$ of layer $l$, weights $\wei{W}^{(l)}\in \mathbb{R}^{n_l \times n_{l-1}}$ of layer $l$, and biases $\wei{b}^{(l)}\in \mathbb{R}^{n_l}$ of layer $l$. The activation function is applied element-wise to its argument. The output of the last layer (and thus the output of the network altogether) is $\wei{x}_{H+1}\in\mathbb{R}^{n_{H+1}}$. Mathematically, an FFNN with $H$ hidden layers establishes a mapping $\mathcal{N}:\mathbb{R}^{n_0}\rightarrow\mathbb{R}^{n_{H+1}}$, $\wei{x}_{H+1} = \mathcal{N}(\wei{x}_{0})$.

    Applying the model of nonlinear viscoelasticity outlined in Sec. \ref{sec:Theory} to compute the stress at each point in time (depending on the strain history) requires implementing Eq. \eqref{eq:reduced_stress_integral_1}. To evaluate this equation for a given strain history, we need to define the following functions: the strain energy $\Psi$ and the reduced relaxation functions $G_r$.  

    \subsection{Strain energy}

    To define the strain energy, we use a CANN \cite{Linka2021} relying on the generalized invariants of the type $\tilde{\mathcal{I}}$. Stresses can be computed by automatic differentiation. The CANN automatically ensures material objectivity, material symmetry, and an energy- and stress-free reference configuration, i.e., $\Psi(\tns{C}=\tns{I})=0$ and $\tns{S}(\tns{C}=\tns{I})=\tns{0}$. The latter is ensured by a term in the strain energy that is continuously adopted during machine learning such that these two conditions remain satisfied. Non-negativeness of the strain energy function, i.e., $\Psi \ge 0$, is ensured by choosing appropriate activation functions and weight constraints in the last two layers of the CANN. In the second-to-last layer, we apply non-negative activation functions $\sigma_{H}:\mathbb{R} \rightarrow \mathbb{R}^{+0}$. In the last layer, we apply a linear activation function $\sigma_{H+1}(\wei{x})=\wei{x}$ and enforce non-negative weights and biases ($\wei{W}^{(H+1)}, \wei{b}^{(H+1)}\ge0$), yielding a non-negative strain energy function. Except for the last layer, where we apply a linear activation function, our default activation function is the softplus function $\sigma_l(x)=\ln(1+\exp(x))$. Apart from the useful property of being positive, the softplus function is a $C^\infty$-continuous function. Hence, the strain energy function, stress tensor, and elasticity tensor are $C^\infty$-continuous functions which is numerically favorable, particularly for implementing the vCANN in FE software.
	
    Note that, if necessary, we can easily guarantee polyconvexity \cite{Ball1976} of the strain energy function when using CANNs. To this end, we enforce non-negative weights in all layers and a non-negative bias in the last layer (but not necessarily in the previous layers \cite{Amos2017}). Then, applying convex, non-decreasing activation functions on all layers renders the network convex \cite{Boyd2004}. We meet this constraint on the activation functions by default since we use linear activation functions in the last layer and softplus activation functions in all other layers. Both activation functions are convex and non-decreasing. Using a neural network with the above features, we only have to ensure that only polyconvex invariants are used in the CANN. In particular, the generalized invariants of the type $\tilde{\mathcal{I}}$ are polyconvex \cite{Ehret2007}. For an overview of other polyconvex invariants, the reader is referred to \cite{Hartmann2003, Schroder2003, Balzani2006}.\\

    \subsection{Reduced relaxation functions} 

    To define the reduced relaxation functions $G_r$ in Eq. \eqref{eq:generalized_Prony}, we must define strain (rate)-dependent relaxation times and relaxation coefficients. To this end, we represent the unknown relation between the strain (rate) and these parameters by FFNNs. We employ separate FFNNs for each relaxation time and coefficient such that each neural network can focus on a particularly simple task.
    The constraints on the reduced relaxation functions, \eqref{eq:Prony_restrictions}, are less severe than those on the strain energy functions. The positivity of the relaxation times and coefficients in Eq. \eqref{eq:Prony_restrictions}$_{2,3}$ is guaranteed by the same methods described above for the strain energy function. The unity constraint on the relaxation coefficients, Eq. \eqref{eq:Prony_restrictions}$_{1}$, is enforced by a custom normalization layer. Apart from that, the functional relations providing the sought parameters are not restricted. Note that using the invariant basis $\mathcal{I}$ as input, the relaxation times and coefficients automatically ensure material objectivity and material symmetry.
	
	The number of Maxwell elements in the generalized Prony series is an important parameter. With sufficiently many Maxwell elements, a Prony series can describe arbitrarily complex viscoelastic materials. Often, materials exhibit numerous different relaxation times \cite{Lakes2009}. A generalized Maxwell model represents each relaxation time by a different Maxwell element. Following these arguments, choosing a large number of Maxwell elements is preferable to represent the relaxation behavior as accurately as possible.
    However, the model complexity, and thus the computational cost, increases with the number of Maxwell elements. Moreover, complex models with many parameters tend to overfit the experimental data, thereby losing the ability to generalize beyond specific given training data. From that perspective, keeping the number of relaxation times and coefficients small is favorable. To balance between an accurate representation of data and a low model complexity, we proceed as follows.
	
	Identifying the relaxation coefficients and times of a generalized Maxwell model is known to be an ill-posed problem \cite{Jalocha2015}. Therefore, in classical approaches, the number of Maxwell elements $N_r$ is determined beforehand and fixed during the parameter identification process \cite{Sharma2020}. In some approaches, the number of Maxwell elements and the relaxation times are determined a priori and fixed during parameter identification to remove ill-posedness \cite{Jalocha2015}. In our approach, we predefine a maximum number of Maxwell elements $N_r^{max}$. During the training, the actual number $N_r$ of Maxwell elements is determined by the vCANN as a part of the learning process within the allowed range $[1;N_r^{max}]$. To avoid unnecessary constraints to the learning process, one may choose relatively large values for $N_r^{max}$, which typically result after the training in $N_r \ll N_r^{max}$. 

    The literature shows that the relaxation times of viscoelastic materials are typically uniformly distributed on a logarithmic scale \cite{Diani2012}. To endow our machine learning architecture with this heuristic prior knowledge, we normalized the output of the $N_r^{max}$ FFNNs that learned the relaxation times by time constants $T_{r\alpha}$, $\alpha=1,2,\ldots, N_r^{max}$. These time constants were uniformly distributed on a logarithmic scale in some range $[T_{min}, T_{max}]$. $T_{min}$ and $T_{max}$ are parameters the user can initially define based on prior knowledge or heuristic expectations. It is important to underline that the normalizing constants $T_{r\alpha}$ are not the relaxation times of our model. The vCANN can, and will in general, learn relaxation times $\tau_{r\alpha}$ (possibly even considerably) differing from the $T_{r\alpha}$. Yet, the $T_{r\alpha}$ provide via the normalization of the output of the FFNNs some bias regarding the expected time scales for the different relaxation times, which can significantly accelerate the training if $[T_{min}, T_{max}]$ is properly chosen. 
 
    Initially, we prescribe the maximum number of Maxwell elements $N_r^{max}$, typically much larger than the actual number $N_r$ required to accurately describe the viscoelastic material. This allows us to gradually eliminate Maxwell elements during training to obtain a sparse model. This approach is similar to the one of \cite{Baumgaertel1989, Baumgaertel1992}, where the number of Maxwell elements was adjusted by merging or removing them during parameter identification to avoid ill-posedness and improve the fit. Likewise, \cite{Rothermel2022} proposed to apply Tikhonov-regularization \cite{Tikhonov1977} to the material parameters and subsequently cluster Maxwell elements with similar relaxation times. We decided to promote sparsity of our model by applying $L_1$ regularization to the relaxation coefficients $g_{r\alpha}$, $\alpha=1,2,\ldots, N_r^{max}$. Using $L_1$ regularization, the optimal value for some relaxation coefficients will be zero, eliminating the corresponding Maxwell elements. In this approach, one uses a penalty parameter $\Lambda$ controlling the sparsity of the model. Choosing $\Lambda=0$ disables regularization, whereas with increasing $\Lambda$, the sparsity of the model increases, too. The penalty parameter $\Lambda$ is a hyperparameter that has to be predefined (and possibly iteratively optimized).
	
    In summary, relaxation times and coefficients, as functions of the invariants $\mathcal{I}$ and the feature vector $\tns{f}$, are learned by individual FFNNs. Scaling of the FFNNs determining the relaxation times by predefined logarithmically uniformly spaced constants introduces a bias in agreement with the literature findings that can help accelerate the training process. $L_1$ regularization on the relaxation coefficients promotes sparse reduced relaxation functions. We implemented the complete vCANN framework using the open-source software library Keras with TensorFlow backend \cite{Chollet2015, Abadi2016}.

    
    \section{Results} \label{sec:results}
    In this section, we apply vCANNs to various data sets. We use synthetic as well as experimental data. In \cite{Linka2021}, we already demonstrated that CANNs could successfully learn the preferred material directions $\vec{l}_{rj}$ and the scalar weight factors $w_{rj}$ in Eq. \eqref{eq:struc_tensor}. Therefore, for simplicity, we herein assume them to be known to focus on this paper's main problem, viscoelastic relaxation. We list the corresponding vCANNs and their hyperparameters in \ref{sec:hyperparameters} for each of the following examples. There, we also provide additional information on the training procedure.

    \subsection{Anisotropic viscoelasticity with synthetic data}\label{sec:res_aniso}
	We created synthetic training data to mimic stress-strain data of viscoelastic soft biological tissues. To this end, we used two hyperelastic constitutive models popular in biomechanics, the Ogden model \cite{Ogden1972}, and the Holzapfel--Gasser--Ogden (HGO) model \cite{Holzapfel2000}. The Ogden model is a phenomenological model for isotropic rubber-like materials and soft biological tissues and is usually formulated in terms of the principal stretches $\lambda_i$, $i=1,2,3$, which are the square roots of the eigenvalues of \tns{C},
	\begin{equation} \label{eq:ogden}
		\Psi_{\mathrm{OG}}(\tns{C}) = \sum_{p=1}^{n} \frac{\mu_p}{\alpha_p} \left( \lambda_1^{\alpha_p} + \lambda_2^{\alpha_p} + \lambda_3^{\alpha_p} -3 \right).
	\end{equation}  
	In Eq. \eqref{eq:ogden}, $n$ is a positive integer, $\mu_p$ and $\alpha_p$ are (constant) material parameters. Many soft biological tissues exhibit stiffening fibers that induce anisotropy. Therefore, the Ogden model is often combined with the HGO model, which adds an anisotropic contribution to the total strain energy function. The strain energy function of the HGO model, with one preferred material direction $\vec{l}$ and structural tensor $\tns{L}= \vec{l}\otimes\vec{l}$, is 
	\begin{equation}
		\Psi_{\mathrm{HGO}}(\tns{C}, \tns{L}) = \begin{cases}
			\frac{k_1}{2k_2}\left\{ \exp \left[ k_2(I_4-1)^2 \right] -1  \right\} &\text{for } I_4 \ge 1, \\[7pt]
			0 &\text{for } I_4 < 1.
		\end{cases} 
	\end{equation}
	$k_1\ge0$ and $k_2>0$ are material parameters, and $I_4=\tns{C}:\tns{L}$. Since $I_4$ represents the squared fiber stretch, $I_4<1$ means compression of the fibers, which are assumed to bear load under tension only. Thus, for $I_4<I$, the anisotropic strain energy and stress contributions are assumed to be zero.

    To produce synthetic data, we used a material model where strain energy was a sum of the Ogden (OG) and HGO strain energy functions, that is, 
 	\begin{equation}
		\Psi = \Psi_{\mathrm{OG}}+ \Psi_{\mathrm{HGO}}.
	\end{equation}
    For the viscous part of the constitutive model used for generating synthetic material data, we assumed strain-dependent relaxation times and coefficients:
	\begin{align}
		\tau^\iso_i(I_1) &= \hat{\tau}_{a,i}^\iso \exp \left( \hat{\tau}_{b,i}^\iso (I_1-3)^2 \right)  , &g^\iso_i(\lambda) &=\hat{g}_{a,i}^\iso \exp \left( \hat{g}_{b,i}^\iso (I_1-3)^2 \right) , &i=&1,2 \label{eq:params_1}\\
		\tau^\ani_1(I_4) &= \hat{\tau}_a^\ani \exp \left( \hat{\tau}_b^\ani (I_4-1)^2 \right), &g^\ani_1(I_4) &= \hat{g}_a^\ani \exp( \hat{g}_b^\ani (I_4-1)^2 ).  \label{eq:params_2}
	\end{align}
    Using the material parameters in Tab. \ref{tab:elastic_params} and Tab. \ref{tab:visco_params} for Eqs. \eqref{eq:params_1} and \eqref{eq:params_2}, we simulated uniaxial cyclic tension-compression experiments with relaxation periods between each tension and compression period. After each complete cycle the stretch rate $\dot{\lambda}$ was changed according to the sequence $\dot{\lambda}=\{ 0.02, 0.03, 0.04, 0.05 \}$ $s^{-1}$. Loading and unloading periods took $t_{move} = 10$ s, respectively. The relaxation periods took $t_{relax} = 60$ s. Thus, a single cycle took $t_{cyc}=160$ s and the total experiment $t_{total}=640$ s. Synthetic training data were generated for different preferred directions, characterized by the acute angle $\varphi$ between the loading and preferred material directions. $\varphi=0^{\circ}$ means that loading direction and preferred direction $\vec{l}$ are parallel, $\varphi=90^{\circ}$ means that both are to orthogonal. The synthetic training data comprised stress data from fictitious materials with four different preferred directions corresponding to $\varphi=\{0, 15, 20, 25\} ^\circ$.
 
	To validate the model, we generated additional synthetic data for a material with the preferred direction $\varphi=10^\circ$, which is not in the training data set. For this material, we simulated two more loading cycles in addition to the above-described loading history, such that the vCANN had to extrapolate the stress response temporally. The stretch rate of the cycles changed according to the sequence $\dot{\lambda}=\{ 0.01, 0.02, 0.03, 0.04, 0.02, 0.05\}$ s$^{-1}$.

\begin{figure}[H]
		\centering
		\includegraphics[width=0.9\linewidth]{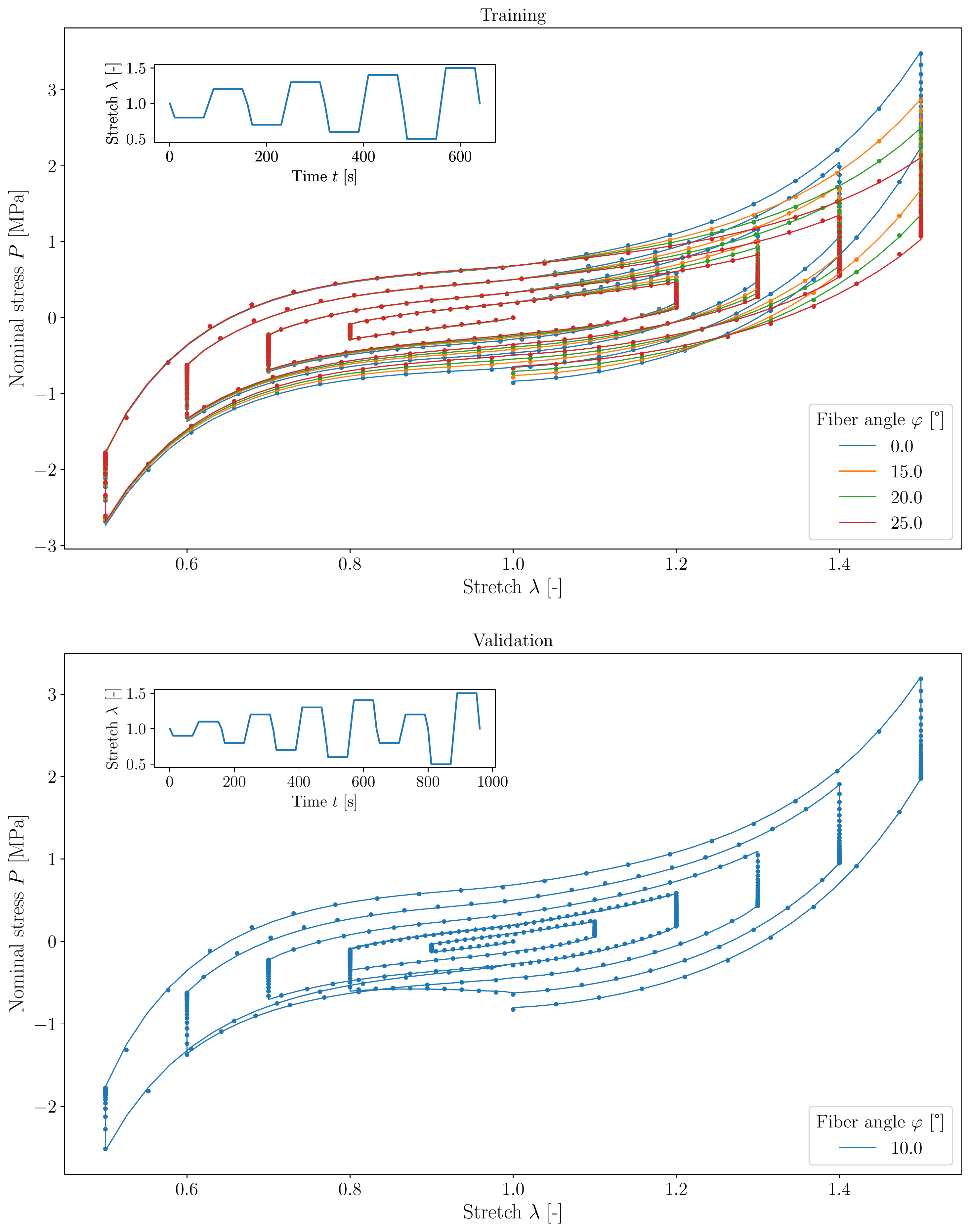}
		\caption{Training (top) and validation (bottom) results. Scatter points represent the synthetic training and validation data, respectively; solid lines represent the vCANN predictions.}
		\label{fig:finalmodelstretch}
	\end{figure}

\begin{figure}[H]
		\centering
		\includegraphics[width=0.9\linewidth]{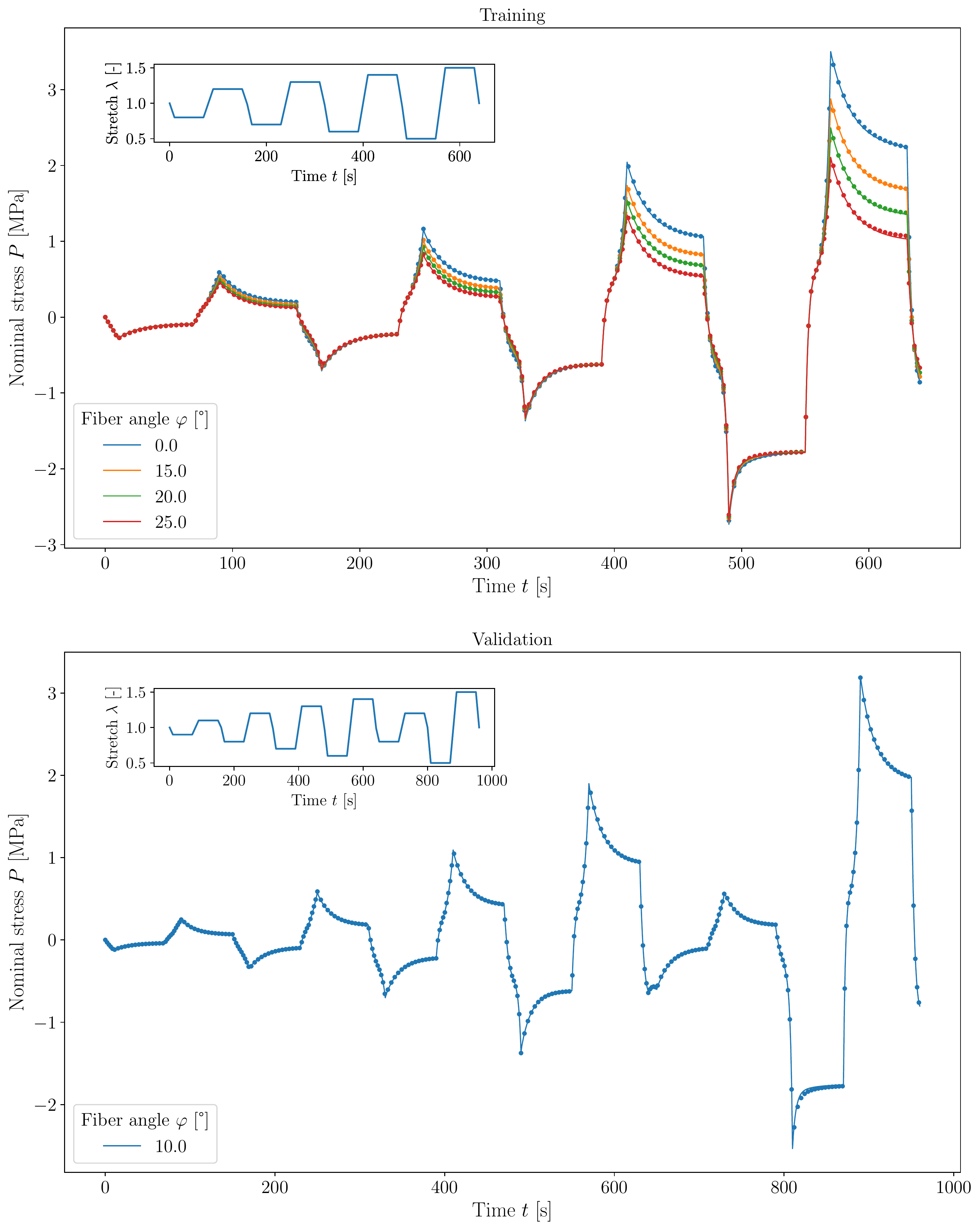}
		\caption{Training (top) and validation (bottom) results. Scatter points represent the synthetic training and validation data; solid lines represent the vCANN predictions.}
		\label{fig:finalmodeltime}
	\end{figure}

    We trained a vCANN with the transversely isotropic structure and hyperparameters given in \ref{sec:hyper_aniso}. Figures \ref{fig:finalmodelstretch} and \ref{fig:finalmodeltime} show that the vCANN learns to replicate the training data almost exactly. In compression, the stress response is similar for all preferred directions since only the isotropic matrix of the composite (Ogden model) bears the load. The vCANN replicates this feature accurately.
    Similarly, the prediction of the validation data set for the unknown preferred direction captures and extrapolates almost perfectly the material response. In particular, the irregular stress response in the time interval $[640, 710]$ s, caused by halving the stretch rate, is predicted precisely.

 
	\subsection{Passive viscoelastic response of the abdominal muscle}\label{sec:res_abdom}
	We reproduced relaxation responses of the leporine rectus abdominis muscle reported in \cite{Calvo2014}. The reduced relaxation function's shape depends on the stretch level, thus exhibiting nonlinear viscoelastic behavior. Classical QLV cannot account for this stretch dependency and would predict the same curve for each stretch level. To represent the stretch-dependent relaxation behavior, \cite{Calvo2014} incorporated stretch-dependent relaxation coefficients and times into a Prony series with one Maxwell element. The authors empirically determined the phenomenological strain dependency of the relaxation coefficients and times. However, their model did not accurately capture the reduced relaxation curves despite utilizing optimization algorithms to fit the material parameters to experimental data (cf. Fig. 10 in \cite{Calvo2014}). This illustrates the limits of human-designed and human-calibrated constitutive models for viscoelastic materials.
	
\begin{figure}[H]
		\centering
		\includegraphics[width=\linewidth]{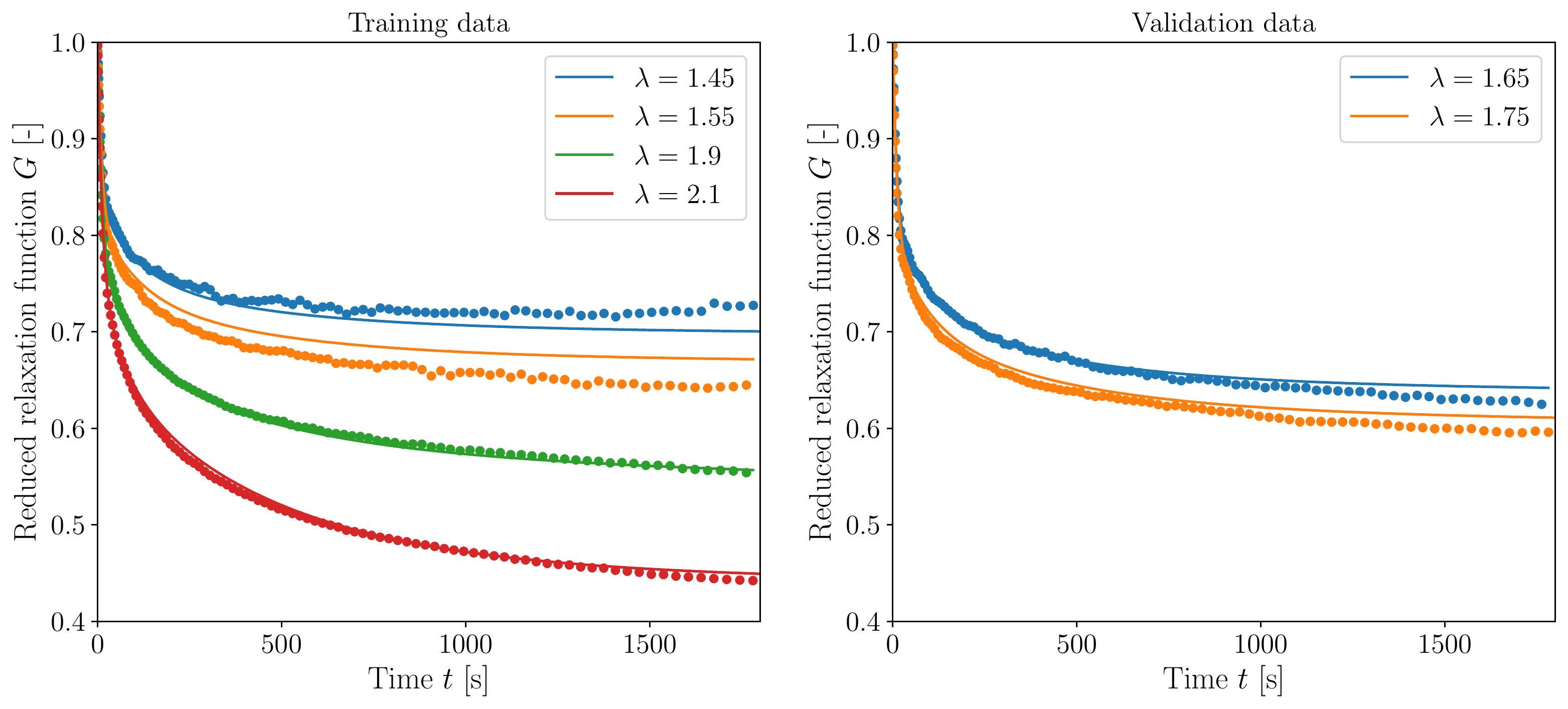}
		\caption{vCANNs can learn to replicate (left) and predict (right) the viscoelastic behavior of abdominal muscle with high accuracy: experimental data from Fig. 4(b) of \cite{Calvo2014} is reproduced by solid circles; solid lines represent the fit of the vCANN.}
		\label{fig:calvo}
	\end{figure}

\begin{figure}[H]
		\centering
		\includegraphics[width=\linewidth]{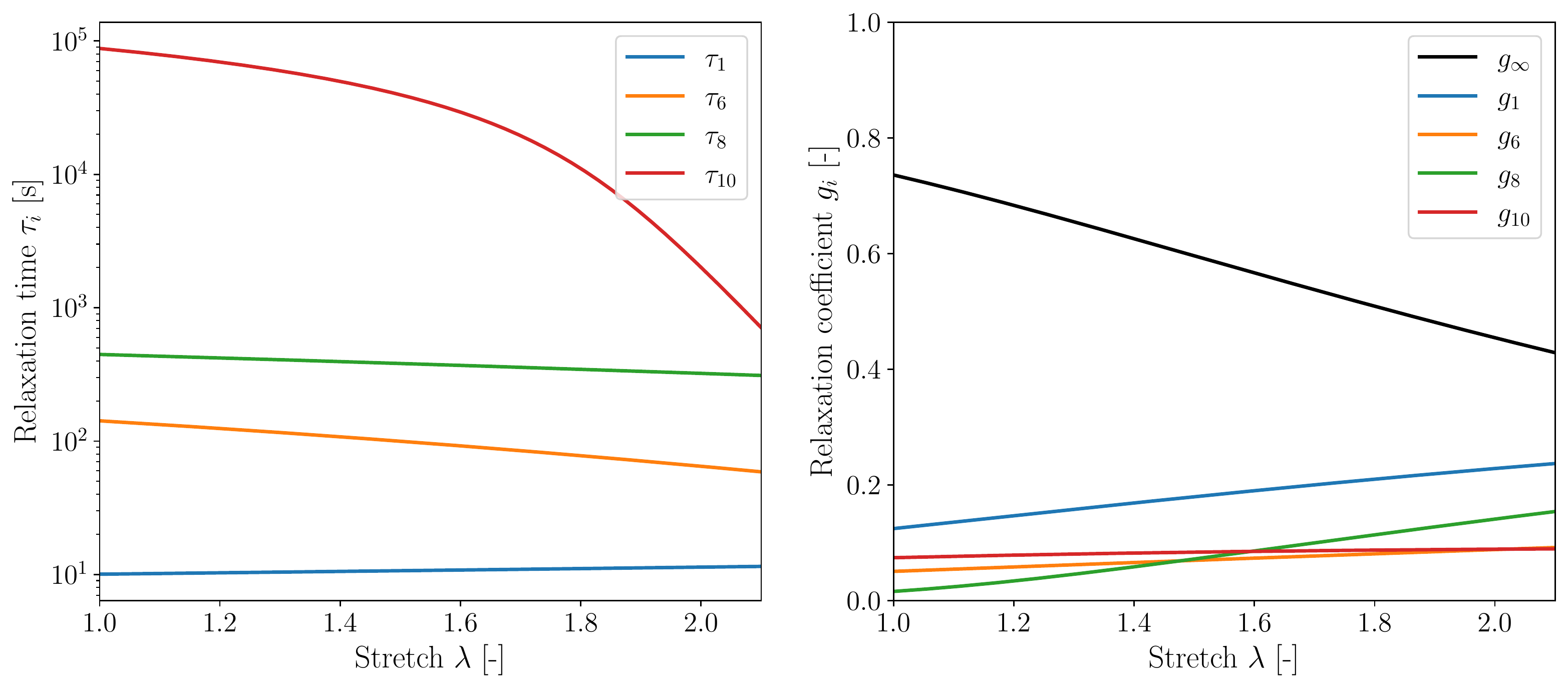}
		\caption{Relaxation times (left) and coefficients (right) learned by the vCANN from the data set of leporine rectus abdominis muscle by \cite{Calvo2014}. Only four of the initial ten Maxwell elements remained after training.}
		\label{fig:prony_calvo}
	\end{figure}

	In contrast, vCANNs capture the relaxation curves with high accuracy (Fig. \ref{fig:calvo}) otherwise only matched by much more complex FNLV models based on the multiplicative split of the deformation gradient, see Fig. 5 in \cite{Latorre2015} for comparison on the same experimental data set. We started the training with $N_{max}=10$ Maxwell elements. Six of them were discarded during training, leaving only the reduce set of parameters plotted in Fig. \ref{fig:prony_calvo}. We provide the trained vCANN structure and the corresponding hyperparameters in \ref{sec:hyper_abdom}.

  
	\subsection{Viscoelastic modeling of VHB 4910}\label{sec:res_vhb4910}
	Very-High-Bond (VHB) 4910 is a soft electro-active polymer (EAP) that exhibits nonlinear viscoelastic behavior and can undergo substantial deformations. VHB 4910 was experimentally studied by \cite{Hossain2012}, using uniaxial loading-unloading tests to characterize the rate-dependent behavior. The tests were conducted for three different stretch rates $\dot{\lambda} = \{ 0.01, 0.03, 0.05\}$ s$^{-1}$ and four different stretch levels $\lambda = \{1.5, 2.0, 2.5, 3.0\}$ (Fig. \ref{fig:vhb_data}). Moreover, the authors conducted a multi-step relaxation test to determine the equilibrium response of the material.
    The constitutive model proposed in \cite{Hossain2012} is based on a multiplicative split of the deformation gradient into an elastic and viscous part. The hyperelastic eight-chain model of \cite{Arruda1993} was chosen to model the elastic part. The material parameters of the elastic part were identified using the data from a multi-step relaxation test. The strain energy function and evolution equation proposed by \cite{Linder2011} were chosen for the viscous part. The viscous material parameters were identified using the loading-unloading data of $\dot{\lambda}=0.01$ s$^{-1}$ and $\dot{\lambda}=0.05$ s$^{-1}$ at a stretch level of $\lambda=3$. The rheological analog model of the constitutive model was a generalized Maxwell model where the number of parallel branches was chosen to be four. 

    A few years later, VHB 4910 was again studied to demonstrate the abilities of a novel advanced microstructurally-informed constitutive model developed in \cite{Zhou2018}. The model relies on advanced knowledge of continuum and statistical mechanics and uses a multiplicative decomposition of the deformation gradient to represent a generalized Maxwell behavior. The elastic material parameters of the model were identified using time-consuming quasi-static tensile tests, and the viscous material parameters using (excluding, however, data with $\lambda=2.5$). The number of Maxwell elements was determined by hand and set to three.
    
\begin{figure}[]
		\centering
		\includegraphics[width=\linewidth]{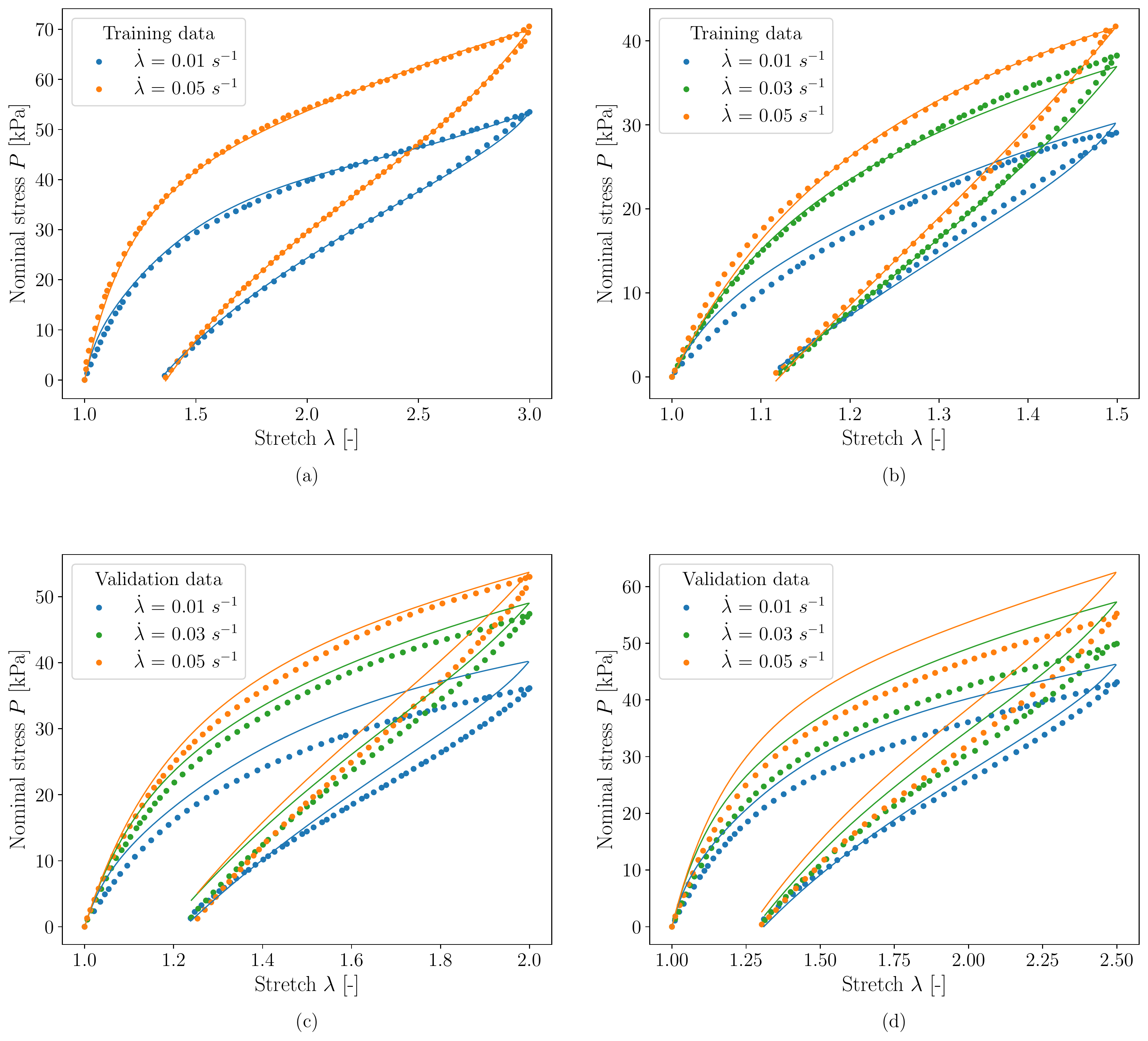}
		\caption{Results of the trained vCANN for the polymer VHB 4910: achieved fitting of training data (a)--(b) and predictive performance on the validation data (c)--(d). Each subfigure shows the loading-unloading stress response for a fixed maximum stretch but different stretch rates. The scatter points represent experimental data on VHB 4910 reproduced from \cite{Hossain2012}; solid lines represent the trained vCANN.}
		\label{fig:vhb}
	\end{figure}

\begin{figure}[]
		\centering
		\includegraphics[width=\linewidth]{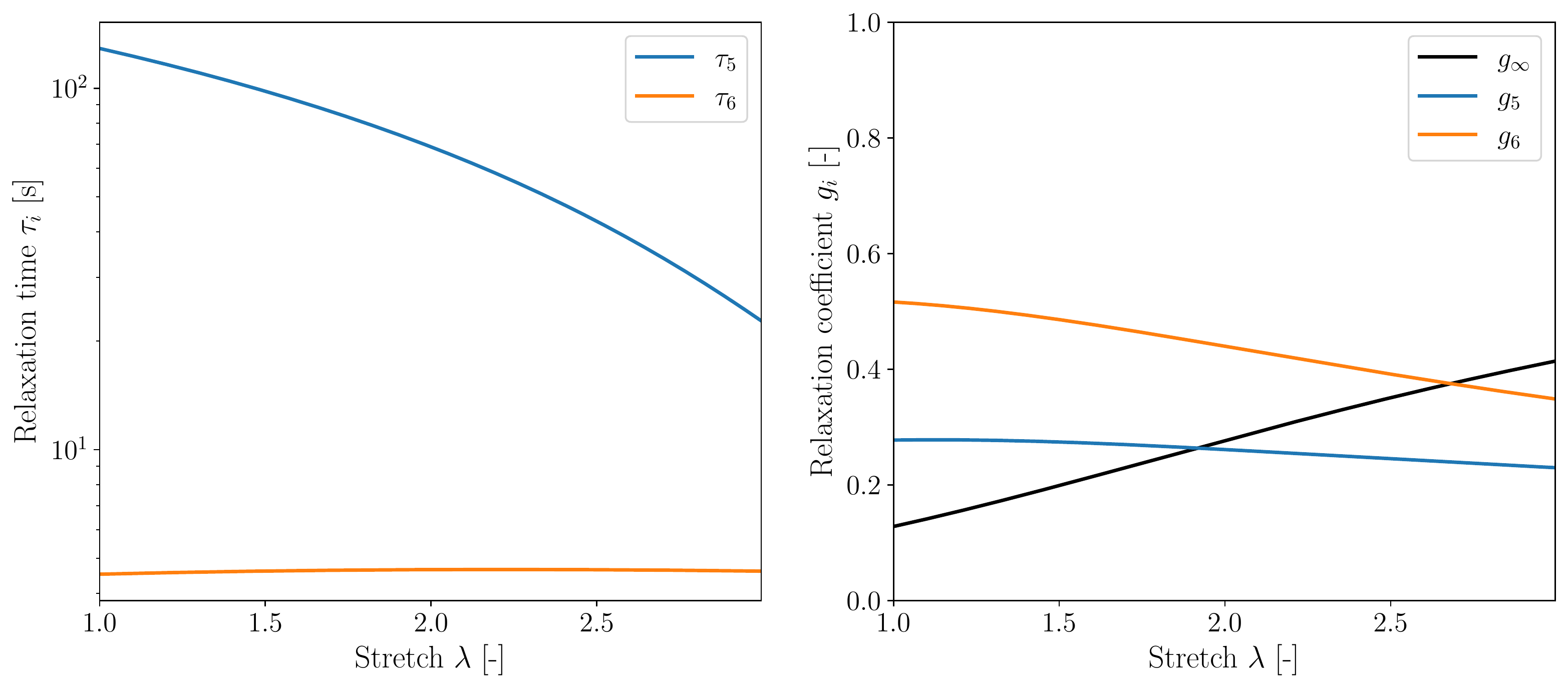}
		\caption{Viscous properties of the vCANN learned from experimental data on VHB 4910 from \cite{Hossain2012}. Only two of the initial 10 Maxwell elements were kept and are necessary to describe the material accurately.}
		\label{fig:vhb_4910_prony}
	\end{figure}

    By contrast, we only used loading-unloading data with $\dot{\lambda}=0.01$ s$^{-1}$ and $\dot{\lambda}=0.05$ s$^{-1}$ at the stretch levels $\lambda=1.5$ and $\lambda=3$ to train the vCANN. Figure \ref{fig:vhb} shows the training and validation results. The fit of both the training and validation is very accurate and at least on par with the one of the FNLV models used in \cite{Hossain2012} (Figs. 9--12), and \cite{Zhou2018} (Fig. 5(b)--(d)). However, we note that the vCANN automatically learned the number of Maxwell elements required to represent the material behavior well. We initialized the vCANN with 10 Maxwell. After training, only two remained, the viscous properties of which we provide in Fig. \ref{fig:vhb_4910_prony}. Moreover, the application of the vCANN did not require advanced expert knowledge and did not require data from particularly sophisticated experiments. These advantages make vCANNs attractive from a practical point of view, in particular in the context of industrial applications. We list details on the trained vCANN structure and the corresponding hyperparameters in \ref{sec:hyper_vhb4910}.

    Remark: The relatively large differences between the experimental data and the vCANN model for $\lambda=2.5$ in Fig. \ref{fig:vhb}(d), which can also be observed for the FNLV model in \cite{Hossain2012}, are likely a result of experimental scatter. The loading paths should be almost identical for a fixed strain rate up to the respective maximum stretches. However, this is not the case, as is highlighted in  Fig. \ref{fig:vhb_data}, which suggests considerable measurement errors in a part of the data, which naturally limited the ability of the vCANN to derive a consistent data-driven model. 
    \FloatBarrier
    
    \subsection{Blast load analysis of Polyvinyl Butyral} \label{sec:res_PVB}
\begin{figure}
		\centering
		\includegraphics[width=\linewidth]{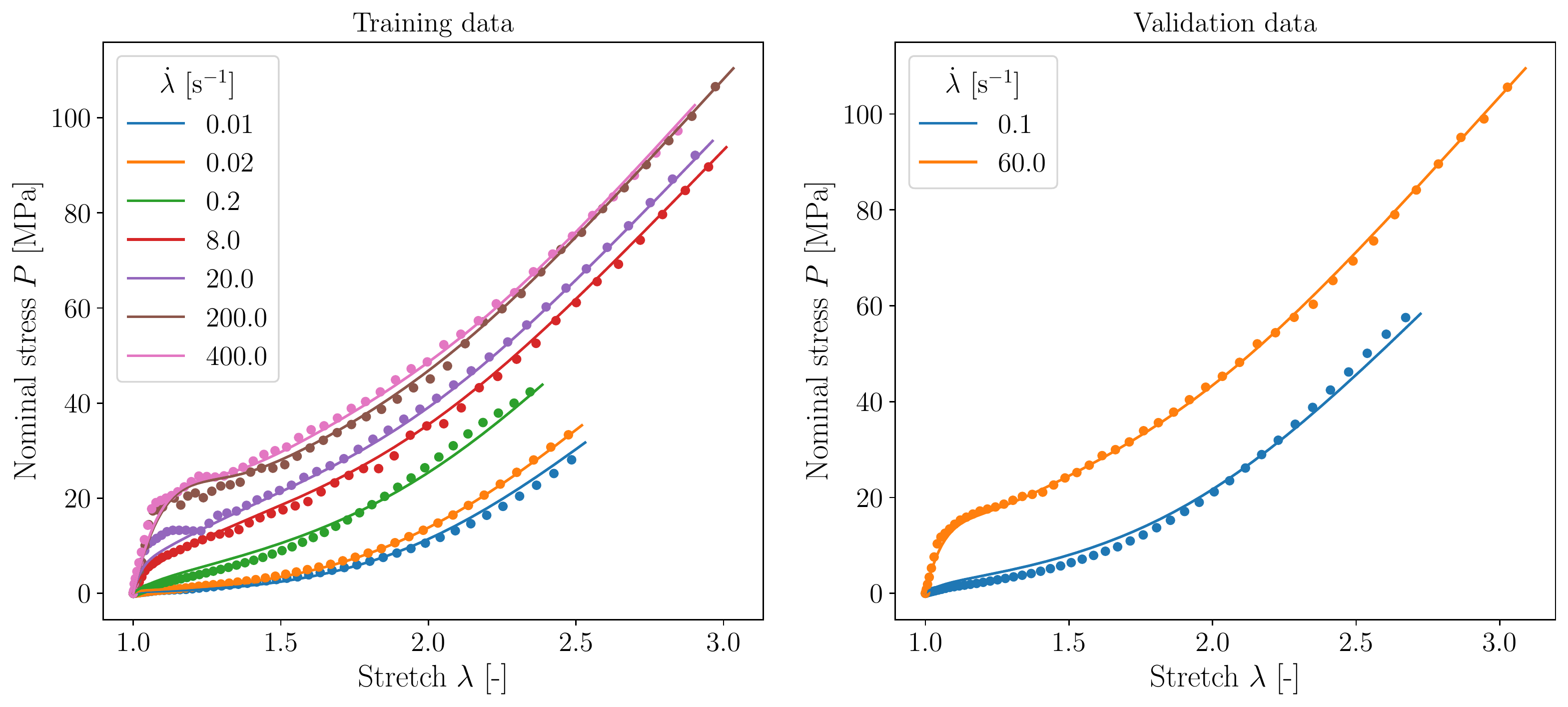}
		\caption{High-stretch rate experiments on PVB. The vCANN accurately describes the constitutive behavior over a wide range of stretch rates for the training (left) and unknown validation data (right). The scatter points represent experimental data reproduced from \cite{Liao2020}; solid lines represent the stress response of the vCANN trained on the complete data set.}
		\label{fig:Linz}
	\end{figure}

	Polyvinyl Butyral (PVB) is a polymer whose primary application is laminated safety glasses. Under heat and pressure, two glass panes are bonded with an interlayer of PVB into a single unit. Under blast loads, the interlayer binds shards of glass, absorbs energy, and mitigates its transfer to the surrounding frame. It is essential to understand the mechanical behavior of PVB to improve the design of laminated glass structures. Large-scale simulations of these structures require simple material models that capture the mechanical behavior over a wide range of strain rates. \cite{DelLinz2016} conducted high-stretch rate experiments on PVB, with stretch rates between $0.01$ s$^{-1}$ and $400$ s$^{-1}$. The viscous properties of PVC likely vary within such a wide range of strain rates. The significant change of the stress-stretch curve's shape above $0.2$ s$^{-1}$ visible in Fig. \ref{fig:Linz} suggests this, too. Ideally, the constitutive model should be able to represent this transition accurately. \cite{DelLinz2016}, proposed an FNLV model and used the strain-dependent viscosity function by \cite{HooFatt2008}. The model describes the experimental data well at high stretch rates, although it cannot accurately resolve the peak stress and subsequent softening at $\lambda\approx1.1-1.2$. At low strain rates, the fit quality is quantitatively unsatisfactory.
    In \cite{DelLinz2016}, the authors also fitted a standard generalized Maxwell model with constant relaxation coefficients and times for comparison. The model comprised six Maxwell elements whose relaxation times were chosen to be uniformly distributed on the logarithmic scale and kept fixed during the parameter identification of the relaxation coefficients. Notably, to account for the broad stretch rate range, two different models had to be used, one for the low stretch rate regime (up to $8$ s$^{-1}$) and the other for the high stretch rate regime ($20$ s$^{-1}$ and above). However, both models could not accurately describe the material behavior in their respective stretch rate regimes.
 
    To account for the rate-dependent viscoelastic properties, we trained the vCANN detailed in \ref{sec:hyper_PVB}. Figure \ref{fig:Linz} shows that the vCANN successfully learned the constitutive behavior over a wide range of stretch rates. Comparing Figs. 12 to 17 in \cite{DelLinz2016} with Fig. \ref{fig:Linz}, reveals that the trained vCANN outperforms the traditional models. In particular, they capture the peak stress and softening in the initial loading phase up to $\lambda\approx1.2$. Importantly, the data-driven nature of our approach apparently provided the flexibility to model the transition between the low and high stretch rate regimes, whereas two different classical models were required to capture the two different regimes. The vCANN did not only learn the constitutive behavior of the training data but also made precise predictions in the low and high stretch regimes for unknown the unknown validation. Remarkably, no advanced expert knowledge was necessary to apply the vCANN, and training the vCANN from scratch took less than 10 minutes on a standard desktop computer. Since the traditional models in \cite{DelLinz2016} were fitted using the entire data set, we also trained the vCANN on the entire data set to ensure a fair comparison.
    \FloatBarrier
    
	\subsection{Thermo-viscoelastic modeling of VHB 4905} \label{sec:res_VHB4905}
\begin{figure}[]
		\centering
		\includegraphics[width=\linewidth]{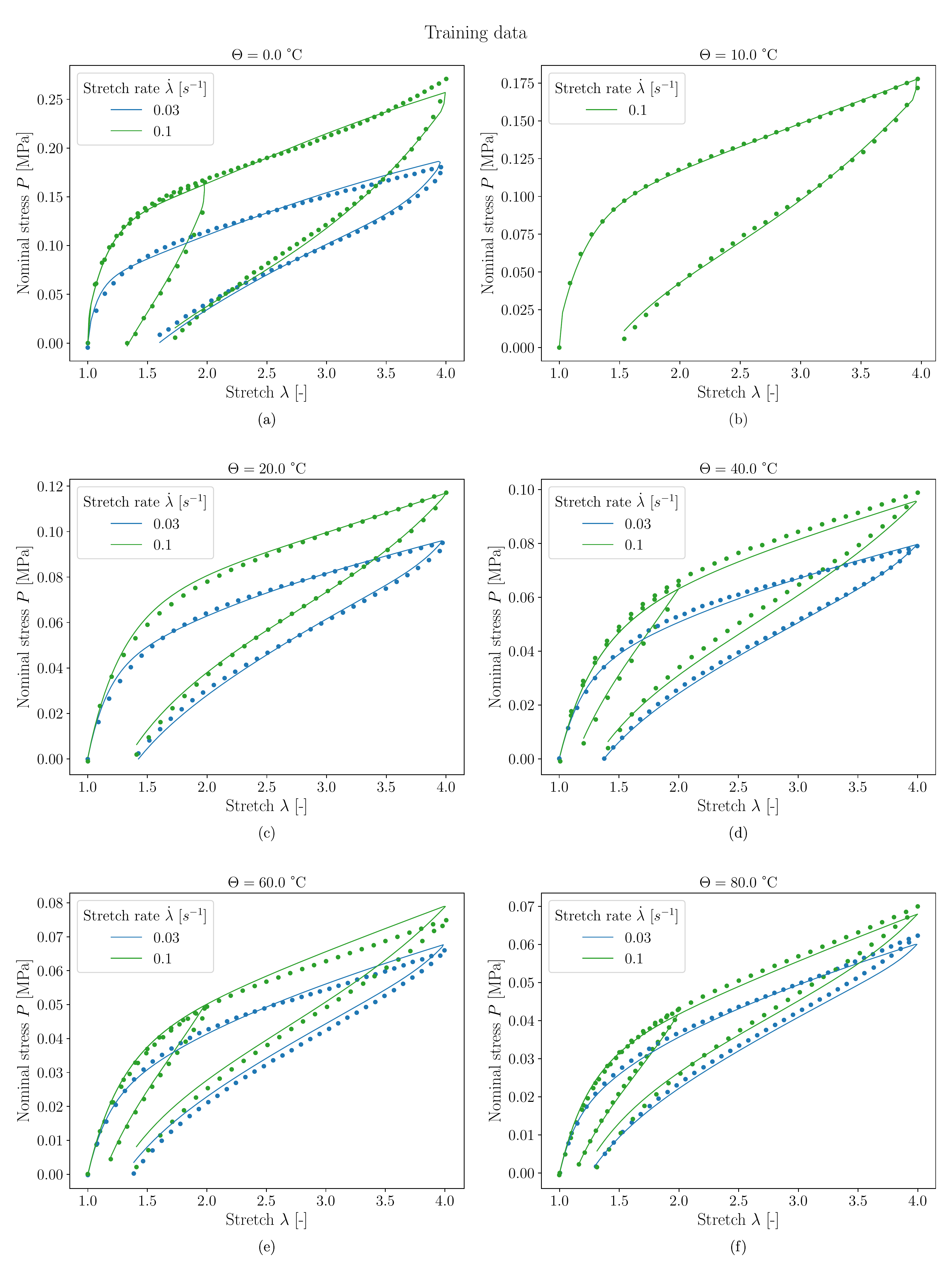}
		\caption{Performance of the vCANN for VHB 4905: achieved fitting of training data. The scatter points represent experimental data reproduced from \cite{Liao2020}; solid lines represent the vCANN performance}
		\label{fig:vhb_4905_training}
	\end{figure}

	Another commercially available EAP is VHB 4905. Like most polymers, VHB 4905 is strongly temperature sensitive. Hence \cite{Liao2020} conducted an extensive experimental study with a wide range of temperatures at different stretch rates and stretch levels. To demonstrate the utility of the feature vector $\tns{f}$ in the vCANN architecture (which is optional and was not yet used in the previous examples), we included the temperature $\Theta$ into the vCANN input as $\tns{f}$. For training we used data of loading-unloading tests with different temperatures $\Theta = \{0, 10 20, 40, 60, 80\}$ [°C] and stretch rates $\dot{\lambda}=\{0.03, 0.1\}$ s$^{-1}$ at the stretch level $\lambda=4$. Additionally, we included in the training set data of tests with $\Theta = \{0, 40, 60, 80\}$ [°C], $\dot{\lambda}= 0.1$ s$^{-1}$ at $\lambda=2$, Fig. \ref{fig:vhb_4905_training}. The strong nonlinear temperature dependence of the stress response is clearly visible by comparing Fig. \ref{fig:vhb_4905_training}(a) and Fig. \ref{fig:vhb_4905_training}(c). In particular, the shape of the stress-stretch curve as well as the stiffness changes between 0°C and 20 °C significantly.
	
\begin{figure}[]
		\centering
		\includegraphics[width=\linewidth]{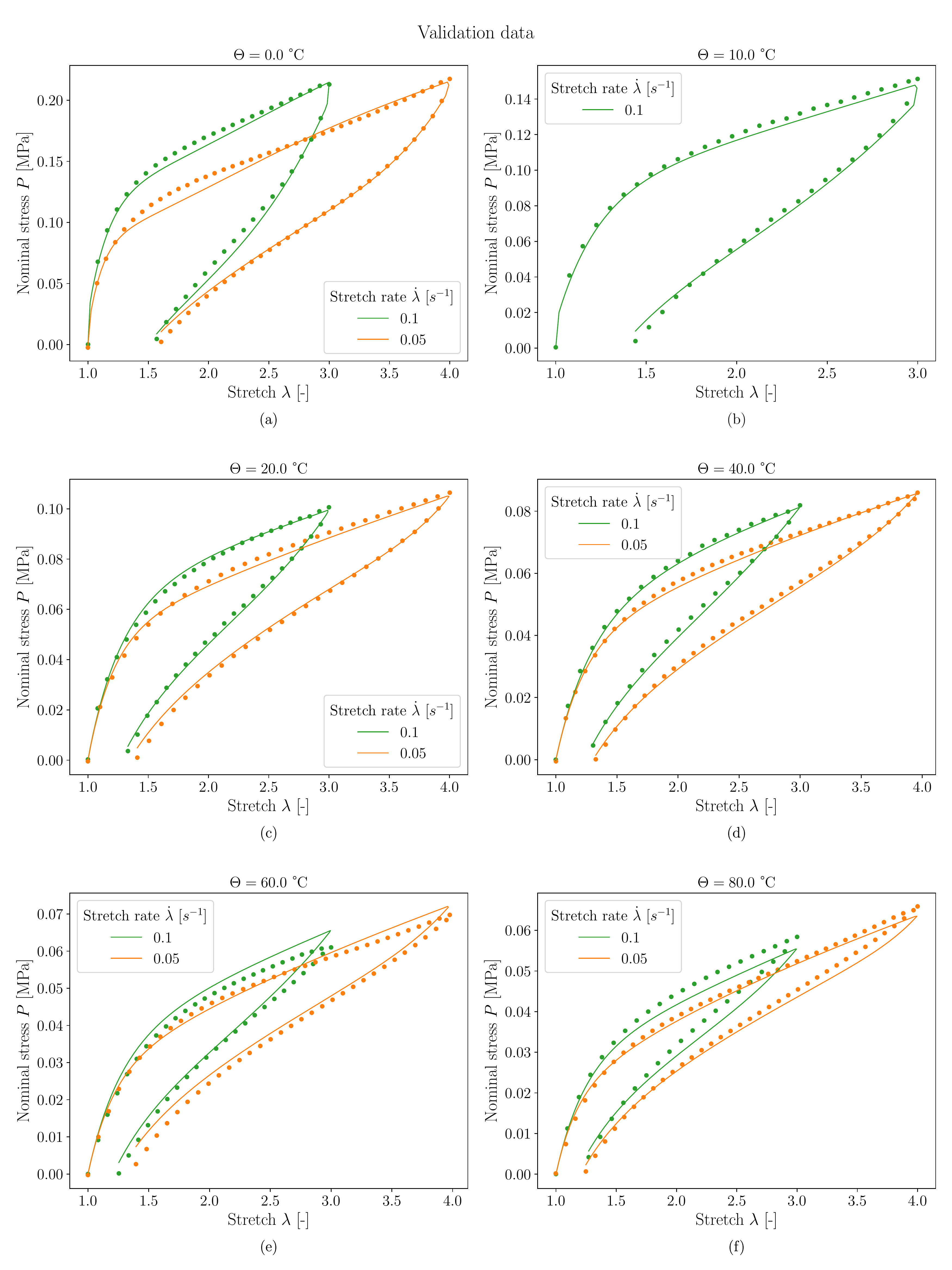}
		\caption{Performance of the vCANN for VHB 4905: predictive performance on validation data. The scatter points represent experimental data reproduced from \cite{Liao2020}; solid lines represent the vCANN performance}
		\label{fig:vhb_4905_validation}
	\end{figure}

	To validate the trained can, we took data from loading-unloading tests with $\Theta = \{0, 10 20, 40, 60, 80\}$ [°C], $\dot{\lambda}=0.03$ s$^{-1}$ at $\lambda=3$. Moreover, we used data from tests with $\Theta = \{0, 10 20, 40, 60, 80\}$ [°C], $\dot{\lambda}=0.05$ s$^{-1}$ at $\lambda=4$ for validation, Figs. \ref{fig:vhb_4905_validation}. Of note, the vCANN had not received any training data with a stretch level $\lambda=3$ nor with a stretch rate $\dot{\lambda}=0.05$ s$^{-1}$. Yet, the trained vCANN was able to predict very well the material behavior for the unknown stretch level and also for the unknown stretch rate. Both is challenging for classical constitutive models and demonstrates the potential of vCANNs. Details on the trained vCANN and its hyperparamters are documented in \ref{sec:hyper_vhb4905}. As seen in \cite{Liao2020}, different sophisticated load protocols are necessary for classical models to calibrate individual parts of the model separately. Although this procedure is possible with vCANNs due to their modularity, they can be trained on a large data set directly, which is much simpler, faster and requires no advanced expert knowledge. 
    \FloatBarrier
    
	\section{Conclusion} \label{sec:conclusion}
    In this paper, we introduced vCANNs, a physics-informed data-driven framework for anisotropic nonlinear viscoelasticity at finite strains. The viscous part is based on a generalized Maxwell model enhanced with nonlinear strain (rate)-dependent relaxation coefficients and times represented by neural networks. The number of Maxwell elements is not determined a priori but adapts automatically during training. Thereby, vCANNs employ $L_1$ regularization on the Maxwell branches to promote a sparse model. In contrast, traditional models usually specify and fix the number of Maxwell branches before calibrating the material parameters, which requires additional, often labor-intense tests.
    vCANNs adopt the computationally very efficient framework of QLV and FLV but generalize these well-established theories to model anisotropic nonlinear viscoelasticity. We demonstrated the ability of vCANNs to learn even challenging viscoelastic behavior of advanced materials by several examples. We also briefly illustrated the ability of vCANNs to process non-mechanical information such as temperature data (or, in other cases also, microstructural or processing data) to predict the behavior of materials under conditions not covered by the training data. We demonstrated that vCANNs could learn the viscoelastic behavior of advanced materials from a database similarly small as the one human experts typically need to calibrate their models. However, vCANNs can learn the material behavior in a fast and fully-automated manner, and their application does not require any expert knowledge. These advantages make vCANNs a favorable tool to support the development of new advanced materials in academia and industry.

    Of note, vCANNs are not only helpful from a practical perspective but can also promote our theoretical understanding. For example, it is often believed that the generalized Maxwell model with strain-dependent material parameters cannot describe strain-dependent relaxation curves accurately \cite{Latorre2015}. Interestingly, the application example on the rectus abdominis muscle presented above demonstrates that vCANNs are very well able to accomplish this. These findings raise the question of whether doubts about the capabilities of generalized Maxwell models are mainly a result of difficulties humans face in their proper calibration instead of fundamental shortcomings of this class of models. In such a way, vCANNs can help us with their automated and highly efficient calibration process to understand the actual capabilities and limits of generalized Maxwell models. Exploring this further may be an exciting avenue for future research.

 
	\section*{Acknowledgements}
	K. P. Abdolazizi and C. J. Cyron gratefully acknowledge financial support from TUHH within the I$^3$-Lab `Modellgestütztes maschinelles Lernen für die Weichgewebsmodellierung in der Medizin'.
    We thank Guang Chen (Department of Mechanical Engineering, University of Connecticut) for sharing parts of his code with us, which is not used in the current version of vCANNs but which was helpful for us to develop ideas.

	\appendix
    \section{Transverse Isotropy}\label{sec:transverse_isotropy}
    To illustrate the proposed constitutive model, we consider a transversely isotropic material. Transversely isotropic materials exhibit one preferred material direction. Material properties remain invariant with respect to rotations about and reflections from the planes orthogonal or parallel to this preferred direction. The preferred direction $\vec{l}_1$ may be interpreted as the direction of a unidirectional family of fibers embedded into some isotropic matrix. We obtain from Eq. \eqref{eq:gen_struc_tensor} the structural tensors
    \begin{align}
        \tns{L}_0 &= \frac{1}{3} \tns{I}, &\tns{L}_1 &= \vec{l}_1 \otimes \vec{l}_1.
    \end{align}
    Setting $\tns{L}_{r1} = \tns{L}_1$, Eq. \eqref{eq:gen_struc_tensor} yields the generalized structural tensors
    \begin{equation}\label{eq:gen_struc_tensor_ti}
        \tilde{\tns{L}}_r = \frac{1}{3} \left( 1- w_{r1} \right) \tns{I} +  w_{r1} \tns{L}_{1}, \quad r = 1,2,\ldots,R.
    \end{equation}
    The generalized structural tensor Eq. \eqref{eq:gen_struc_tensor_ti} describes a transverely isotropic fiber dispersion with rotational symmetry around a mean fiber direction aligned with $\vec{l}_1$ \cite{Gasser2006}. Unidirectional alignment requires the uncoupling of the two contributions $\tns{I}$ and $\tns{L}_{1}$. Hence, setting $R=2$, $w_{11}=w_{20}=0$, and $w_{10}=w_{21}=1$ results in
    \begin{align}
        \tilde{\tns{L}}_1 &= \frac{1}{3}\tns{I}, &\tilde{\tns{L}}_2&=\tns{L}_{1}.
    \end{align}
    With Eqs. \eqref{eq:generalized_invariants_1} and \eqref{eq:generalized_invariants_2}, the generalized invariants are
    \begin{align}
        \tilde{I}_1&=\frac{1}{3} \tr\left( \tns{C} \right) , &\tilde{J}_1&= \frac{1}{3} \tr \left( \cof \tns{C} \right) , &\tilde{I}_2&=\tr\left( \tns{C}\tns{L}_1 \right) , &\tilde{J}_2&=\tr \left(\left( \cof \tns{C} \right) \tns{L}_1 \right) &\mathrm{III}_{\tns{C}}&=\det\tns{C}=1,
    \end{align}
    and
    \begin{align}
        \tilde{\dot{I}}_1&=\frac{1}{3} \tr\left( \dot{\tns{C}} \right) , &\tilde{\dot{J}}_1&= \frac{1}{3} \tr \left( \cof \dot{\tns{C}} \right) , &\tilde{\dot{I}}_2&=\tr\left( \dot{\tns{C}}\tns{L}_1 \right) , &\tilde{\dot{J}}_2&=\tr \left(\left( \cof \dot{\tns{C}} \right) \tns{L}_1 \right), &\mathrm{III}_{\dot{\tns{C}}}&=\det\dot{\tns{C}},
    \end{align}
    such that
    \begin{align}
        \tilde{\mathcal{I}} &= \left\{  \tilde{I}_1, \tilde{J}_1, \tilde{I}_2, \tilde{J}_2 \right\}, &\tilde{\dot{\mathcal{I}}} &= \left\{  \tilde{\dot{I}}_1, \tilde{\dot{J}}_1, \tilde{\dot{I}}_2, \tilde{\dot{J}}_2, \mathrm{III}_{\dot{\tns{C}}} \right\}, &\mathcal{I} &= \tilde{\mathcal{I}} \cup \tilde{\dot{\mathcal{I}}}.
    \end{align}
    According to Eq. \eqref{eq:2_PK_general}, the instantaneous elastic 2. Piola--Kirchhoff stress of a transversely isotropic material with unidirectional fiber alignment can be computed by differentiating the strain energy function
    \begin{equation}\label{eq:SEF_ti}
        \Psi = \Psi\left( \mathcal{I}, \tns{f}\right)
    \end{equation}
    with respect to $\tns{C}$, giving
    \begin{equation}\label{eq:stress_ti}
        \tns{S}^e = - p\tns{C}^{-1} + 2 \left( \partialder{\Psi}{\tilde{I}_1} \tns{I} - \partialder{\Psi}{\tilde{J}_1} \tns{C}^{-2} \right)
        + 2 \left( \partialder{\Psi}{\tilde{I}_2} \tns{L}_1 - \partialder{\Psi}{\tilde{J}_2} \tns{C}^{-1} \tns{L}_1 \tns{C}^{-1} \right).
    \end{equation} 
    The reduced relaxation functions Eq. \eqref{eq:reduced_relax_2} simplify to
    \begin{equation}\label{eq:reduced_relax_ti}
         G_{r}= G_r \left( t; \mathcal{I},  \tns{f} \right), \quad r=1,2.
    \end{equation}
    Within the proposed framework of anisotropic nonlinear viscoelasticity, Eqs. \eqref{eq:stress_ti} and \eqref{eq:reduced_relax_ti} constitute the most general expressions for the stress and reduced relaxation functions of a transversely isotropic material with unidirectional fiber alignment. For practical applications, it is often useful to uncouple $\Psi$ and $G_r$ with respect to the generalized structural tensors:
    \begin{equation}\label{eq:SEF_ti_uncoupled}
        \Psi = \Psi_1(\tilde{I}_1, \tilde{J}_1, \tns{f}) + \Psi_2(\tilde{I}_2, \tilde{J}_2, \tns{f}),
    \end{equation}
    \begin{align}\label{eq:reduced_relax_ti_uncoupled}
         G_{1} &= G_1 \left( t; \tilde{I}_1, \tilde{J}_1, \tilde{\dot{I}}_1, \tilde{\dot{J}}_1, \mathrm{III}_{\dot{\tns{C}}}, \tns{f} \right),
         &G_{2}&= G_2 \left( t; \tilde{I}_2, \tilde{J}_2, \tilde{\dot{I}}_2, \tilde{\dot{J}}_2, \mathrm{III}_{\dot{\tns{C}}}, \tns{f} \right).
    \end{align}
    Uncoupling can significantly accelerate the training process of vCANNs. We can identify $\Psi_1$ and $G_1$ with the isotropic strain energy function and reduced relaxation function, respectively. By contrast, $\Psi_2$ and $G_2$ represent the anisotropic strain energy function and reduced relaxation function.
    This example illustrates our proposed framework's versatility and that it includes important classes of anisotropy as special cases.
    
	\section{Numerical time integration}\label{sec:numericalintegration}
    In this section, we provide the derivation of the numerical time-stepping scheme used within the vCANN framework. We are interested in computing the viscous overstresses (Eq. \eqref{eq:integral_1}
    \begin{equation}
        \tns{Q}_{r\alpha} = \int_{-\infty}^t g_{r\alpha}(\mathcal{I}, \tns{f}) \exp \left( {-\frac{t-s}{\tau_{r\alpha}(\mathcal{I}, \tns{f})}} \right) \dot{\tns{S}}_r^e \;\mathrm{d}s.
    \end{equation}
    To this end, we recall the evolution equation of a single Maxwell branch with a strain (rate)-dependent relaxation coefficient and time.
	The evolution of the viscous overstress $\tns{Q}_{r\alpha}$ is governed by the linear ODE of first order with variable coefficients and with some known but otherwise arbitrary initial value $\tns{Q}^n_{r\alpha}$ at an arbitrary time point $t^n$,
	\begin{equation}\label{eq:ode_1}
		\dot{\tns{Q}}_{r\alpha} + \frac{\tns{Q}_{r\alpha}}{\tau_{r\alpha}(\mathcal{I}, \tns{f})} = g_{r\alpha}(\mathcal{I},\tns{f}) \, \dot{\tns{S}}_r^e, \quad \tns{Q}^n_{r\alpha} = \tns{Q}_{r\alpha}(t^n).
	\end{equation}
	This equation can be solved by a time-stepping scheme after discretizing time into a number of time points $t^i$. Consider the small time interval $[t^n, t^{n+1}]$ between time points $t^n$ and $t^{n+1}$
    with time step size $\Delta t = t^{n+1} - t^n$. For sufficiently small intervals (and assuming a sufficiently smooth problem) $\Delta t$, $\tau_{r\alpha}$ and $g_{r\alpha}$ can through the whole time interval be approximated by the average values of its beginning and end point:
    \begin{align}\label{equation:appA:averagetaug}
        \bar{\tau}_{r\alpha} &= \frac{\left( \tau_{r\alpha} \right)^{n+1} + \left( \tau_{r\alpha} \right)^{n}}{2}, \quad \quad &\bar{g}_{r\alpha} &= \frac{\left( g_{r\alpha} \right)^{n+1} + \left( g_{r\alpha} \right)^n}{2}.
    \end{align}
	In a displacement-driven setting, $\bar{\tau}_{r\alpha}$ and $\bar{g}_{r\alpha}$ are known since they depend on the prescribed deformation (rate) at the considered times. With the approximation Eq. \eqref{equation:appA:averagetaug}, Eq. \eqref{eq:ode_1} becomes a linear ODE of first order with constant coefficients and with some known initial value:
	\begin{equation}\label{eq:linearized_ode_appx}
		\dot{\tns{Q}}_{r\alpha} + \frac{\tns{Q}_{r\alpha}}{\bar{\tau}_{r\alpha}} = \bar{g}_{r\alpha} \dot{\tns{S}}_r^e, \quad \tns{Q}^n_{r\alpha} = \tns{Q}_{r\alpha}(t^n).
	\end{equation}
	Multiplying both sides of Eq. \eqref{eq:linearized_ode_appx} by the integrating factor $\exp(t/\bar{\tau}_{r\alpha})$ and applying the product rule gives
	\begin{equation}\label{eq:ode_3}
		\frac{\mathrm{d}}{\mathrm{d}t}\left[ \tns{Q}_{r\alpha} \exp\left(\frac{t}{\bar{\tau}_{r\alpha}}\right) \right]  = \bar{g}_{r\alpha} \dot{\tns{S}}_r^e \exp\left(\frac{t}{\bar{\tau}_{r\alpha}}\right).
	\end{equation}
	Integrating Eq. \eqref{eq:ode_3} from $t^n$ to $t^{n+1}$ yields
	\begin{equation}
		\exp\left(\frac{t^{n+1}}{\bar{\tau}_{r\alpha}}\right) \tns{Q}_{r\alpha}^{n+1} - \exp\left(\frac{t^n}{\bar{\tau}_{r\alpha}}\right) \tns{Q}_{r\alpha}^{n}
		= \int_{t^n}^{{t^{n+1}}} \exp\left(\frac{t}{\bar{\tau}_{r\alpha}}\right) \bar{g}_{r\alpha} \dot{\tns{S}}_r^e \, \mathrm{d}t
	\end{equation}
	which can subsequently be solved for 
	\begin{align}		
		\tns{Q}_{r\alpha}^{n+1} &= \exp\left(\frac{t^n}{\bar{\tau}_{r\alpha}}\right) \exp\left(-\frac{t^{n+1}}{\bar{\tau}_{r\alpha}}\right) \tns{Q}_{r\alpha}^{n} + \int_{t^n}^{{t^{n+1}}} \exp\left(\frac{t}{\bar{\tau}_{r\alpha}}\right) \exp\left(-\frac{t^{n+1}}{\bar{\tau}_{r\alpha}}\right) \bar{g}_{r\alpha} \dot{\tns{S}}_r^e \, \mathrm{d}t\\
		&= \exp\left(-\frac{\Delta t}{\bar{\tau}_{r\alpha}}\right) \tns{Q}_{r\alpha}^{n} + \int_{t^n}^{{t^{n+1}}} \exp \left( -\frac{t^{n+1}-t}{\bar{\tau}_{r\alpha}} \right) \bar{g}_{r\alpha} \dot{\tns{S}}_r^e \, \mathrm{d}t \label{eq:integral_3}\\
		&\approx \exp\left(-\frac{\Delta t}{\bar{\tau}_{r\alpha}}\right) \tns{Q}_{r\alpha}^{n} + \exp \left( -\frac{\Delta t}{2\bar{\tau}_{r\alpha}} \right) \bar{g}_{r\alpha} \int_{t^n}^{{t^{n+1}}} \dot{\tns{S}}_r^e \, \mathrm{d}t \\
		&= \exp\left(-\frac{\Delta t}{\bar{\tau}_{r\alpha}}\right) \tns{Q}_{r\alpha}^{n} + \exp \left( -\frac{\Delta t}{2\bar{\tau}_{r\alpha}} \right) \bar{g}_{r\alpha} \left[ (\tns{S}_r^e)^{n+1} - (\tns{S}_r^e)^n \right] \label{eq:update_1}
	\end{align}
	which yields a recurrence update formula for the viscous overstress at time $t^{n+1}$, given we know the state at time $t^n$. We used the mid-point rule on the integral in Eq. \eqref{eq:integral_3} approximating the time variable $t$ by $(t^{n+1}-t^n)/2$.\\
	
	An alternative update formula for the overstress $\tns{Q}_{r\alpha}^{n+1}$ can be obtained by approximating $\dot{\tns{S}}_r^e \approx \frac{(\tns{S}_r^e)^{n+1} - (\tns{S}_r^e)^{n}}{\Delta t}$ directly in Eq. \eqref{eq:integral_3}, which leads to
	\begin{align}
		\tns{Q}_{r\alpha}^{n+1} &= \exp\left(-\frac{\Delta t}{\bar{\tau}_{r\alpha}}\right) \tns{Q}_{r\alpha}^{n} + \bar{g}_{r\alpha} \frac{(\tns{S}_r^e)^{n+1} - (\tns{S}_r^e)^{n}}{\Delta t} \int_{t^n}^{{t^{n+1}}} \exp \left( -\frac{t^{n+1}-t}{\bar{\tau}_{r\alpha}} \right) \mathrm{d}t \\
		&= \exp\left(-\frac{\Delta t}{\bar{\tau}_{r\alpha}}\right) \tns{Q}_{r\alpha}^{n} + \frac{\bar{g}_{r\alpha} \bar{\tau}_{r\alpha}}{\Delta t} \left[ 1 - \exp \left( -\frac{\Delta t}{\bar{\tau}_{r\alpha}} \right) \right] \left[ (\tns{S}_r^e)^{n+1} - (\tns{S}_r^e)^{n} \right], \label{eq:update_2}
	\end{align}
	giving an update formula for the overstress at time $t^{n+1}$. The two update formulae \eqref{eq:update_1} and \eqref{eq:update_2} are very similar to common recurrence formulae for the stress update found in the literature \cite{Simo1987, Holzapfel1996, Simo1998}. The major difference to most formulae reported in the literature is that $\bar{\tau}_{r\alpha}$ and $\bar{g}_{r\alpha}$ are not constant but change each time step. In essence, in a discrete time stepping scheme, one solves a different QLV problem in each time step.

	The algorithmic linearization of the stress tensor $\tns{S}^{n+1}$ is essential for solving nonlinear boundary problems. With Eq. \eqref{eq:integral_1}, we obtain an update rule for the stress tensor $\tns{S}^{n+1}$ at time point $t^{n+1}$:
	\begin{equation}
		\tns{S}^{n+1} = -\left( p\tns{C}^{-1} \right)^{n+1} + \sum_{r=1}^R \left( (\tns{S}_r^{\infty})^{n+1} + \sum_{\alpha=1}^{N_r} \tns{Q}_{r\alpha}^{n+1} \right).
	\end{equation}

	\section{Structure learning block}\label{sec:struc_learn}
        \begin{figure}[H]
		\centering
		\includegraphics[width=\linewidth]{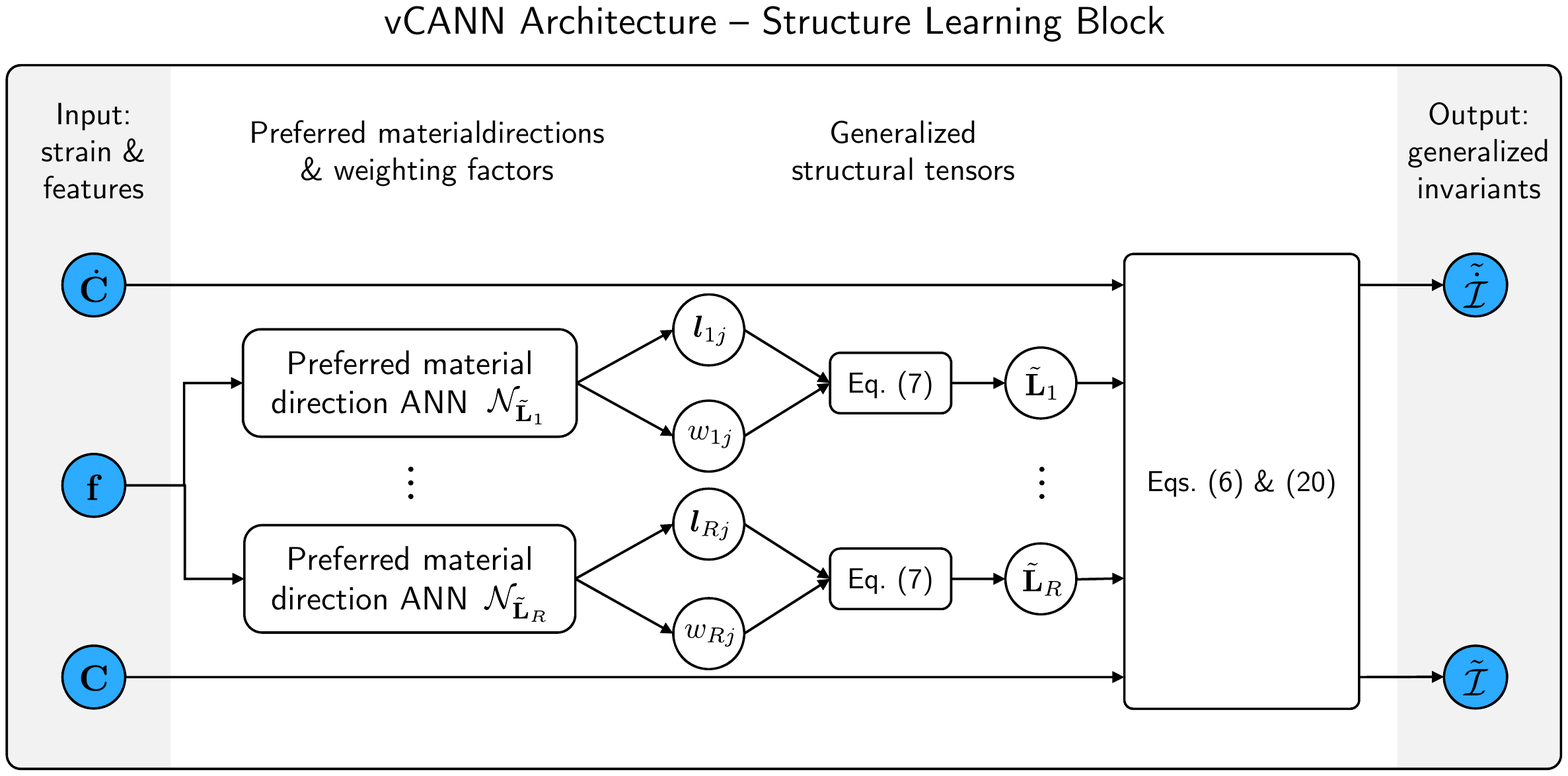}
		\caption{Schematic illustration of the structure learning block: The deformation (rate) tensors $\tns{C}$, $\dot{\tns{C}}$, and the feature vector $\tns{f}$ serve as input to the structure learning block. The preferred material directions $\vec{l}_{rj}$ and scalar weights $w_{rj}$ are learnt from the feature vector $\tns{f}$ within dedicated neural networks $\mathcal{N}_{\tilde{\tns{L}}_r}$. Inserting the outputs of $\mathcal{N}_{\tilde{\tns{L}}_r}$ in Eq. \eqref{eq:gen_struc_tensor} yields the generalized structural tensors $\tilde{\tns{L}}_r$. Together with the deformation (rate) tensors $\tns{C}$ and $\dot{\tns{C}}$, we obtain from Eqs. \eqref{eq:generalized_invariants_1} and \eqref{eq:generalized_invariants_2} the generalized invariants $\tilde{\mathcal{I}}$ and $\tilde{\dot{\mathcal{I}}}$, respectively. In turn, the generalized invariants $\tilde{\mathcal{I}}$ and $\tilde{\dot{\mathcal{I}}}$ themselves serve as input to the main part of the vCANN in Fig. \ref{fig:architecture}.}
		\label{fig:struc_learn}
	\end{figure}
 
	\section{Material parameters for the Ogden and HGO model} \label{sec:mat_params}
	\begin{table}[H]
		\centering	
		\parbox{\linewidth}{
			\centering	
			\begin{tabular*}{0.25\textwidth}{@{\extracolsep{\fill}}cccc}
				\hline
				$\mu_1$ & $\alpha_1$ & $k_1$ & $k_2$ \\
				\hline
				0.3 & 3.7 & 0.3& 0.4\\
			\end{tabular*}
			\caption{Isotropic and anisotropic elastic material parameters used for synthetic data generation}
			\label{tab:elastic_params}
		}
		\vspace*{1.cm}
		
		\parbox{\linewidth}{	
			\centering
			\begin{tabular*}{0.9\textwidth}{@{\extracolsep{\fill}}cccccccccccc}
				\hline
				$\hat{\tau}^\iso_{a,1}$ & $\hat{\tau}^\iso_{b,1}$ & $\hat{g}^\iso_{a,1}$ & $\hat{g}^\iso_{b,1}$ & $\hat{\tau}^\iso_{a,2}$ & $\hat{\tau}^\iso_{b,2}$ & $\hat{g}^\iso_{a,2}$ & $\hat{g}^\iso_{b,2}$   & $\hat{\tau}^\ani_{a,2}$ & $\hat{\tau}^\ani_{b,2}$ & $\hat{g}^\ani_{a,2}$ & $\hat{g}^\ani_{b,2}$\\
				\hline
				20.0&  -7.0&  0.4&  -2.8&  1.0&  4.0&  0.1&  -2.8 & 10.0& 0.7 & 0.8& -1.1\\
			\end{tabular*}
			\caption{Isotropic and anisotropic viscoelastic material parameters used for synthetic data generation}
			\label{tab:visco_params}
		}
	\end{table}

    \newcommand*{\tabindent}{ \hspace{5mm}}
    \section{Model training and hyperparameters}\label{sec:hyperparameters}
    In the following we list the vCANNs trained in Sec. \ref{sec:results} together with their hyperparameters. The selection of activation functions is discussed in Sec. \ref{sec:architecture}. For training the vCANNs, we used the mean squared error (MSE) between the actual stress response and the one estimated by the vCANN as the loss function. The gradients of the loss with respect to model parameters are calculated by the backpropagation algorithm using automatic differentiation. The training was terminated based on early stopping. All weights and biases were initialized with Glorot/Xavier uniform initializer and zeros, respectively. No regularization (weight decay) was applied to the weights and biases during training. No dropout layers were used. All vCANNs were trained with Adam optimizer ($\beta_1=0.9$, $\beta_2=0.999$, $\varepsilon=10^{-7}$).
    
    \subsection{Anisotropic viscoelasticity with synthetic data, Sec. \ref{sec:res_aniso}}\label{sec:hyper_aniso}
    The material is transversely isotropic, exhibits strain-dependent but no strain rate-dependent viscous effects, and has no notable features ($\tns{f}=\tns{0}$). Thus, according to Eqs. \eqref{eq:SEF_ti_uncoupled} and \eqref{eq:reduced_relax_ti_uncoupled}, the vCANNs is given by
    \begin{align}
        \Psi &= \Psi_1(\tilde{I}_1, \tilde{J}_1) + \Psi_2(\tilde{I}_2, \tilde{J}_2), &G_1 &= G_1 \left( t; \tilde{I}_1, \tilde{J}_1\right), &G_2&= G_2 \left( t; \tilde{I}_2, \tilde{J}_2 \right).
    \end{align}
    \begin{table}[H]
        \centering
        \renewcommand{\arraystretch}{1.2}
        \begin{tabular}{ll}
            \toprule
            Hyperparameter & Value \\
            \midrule
            \emph{General} & \\
    		\tabindent Learning rate &  $0.001$\\
    		\tabindent Sparsity penalty parameter $\Lambda$ &  $0.001$ \\            
            \midrule
            \emph{Instantaneous elastic stress (CANN)} & \\
            \tabindent Convex & Yes\\
            \tabindent Number of neurons per hidden layer ($\Psi_1$) &  $\{32,32,32\}$ \\
            \tabindent Number of neurons per hidden layer ($\Psi_2$) &  $\{32,32,32\}$ \\
            \midrule
            \emph{Reduced relaxation functions} & \\
            \tabindent Maximal number of Maxwell elements $N_1^{max}$ & 5\\
            \tabindent Maximal number of Maxwell elements $N_2^{max}$ & 5\\
            \tabindent Time normalization $[T_{min}, T_{max}]$ & $[10^{-2}, 10^{3}]$ s\\
            \tabindent Number of neurons per hidden layer of $\mathcal{N}_{\tau_{1\alpha}}$ & $\{ 32, 32, 16 \}$ \\
            \tabindent Number of neurons per hidden layer of $\mathcal{N}_{g_{1\alpha}}$ & $\{ 32, 32, 16 \}$ \\
            \tabindent Number of neurons per hidden layer of $\mathcal{N}_{\tau_{2\alpha}}$ & $\{ 32, 32, 16 \}$ \\
            \tabindent Number of neurons per hidden layer of $\mathcal{N}_{g_{2\alpha}}$ & $\{ 32, 32, 16 \}$ \\
            \bottomrule
        \end{tabular}
        \caption{Hyperparameters of the vCANN from Sec. \ref{sec:res_aniso}}
        \label{tab:hyper_aniso}
    \end{table}

    \subsection{Passive viscoelastic response of the abdominal muscle, Sec. \ref{sec:res_abdom}}\label{sec:hyper_abdom}
    The material is isotropic, exhibits strain-dependent but no strain rate-dependent viscous effects, and has no notable features ($\tns{f}=\tns{0}$). Thus, according to Eqs. \eqref{eq:SEF_ti_uncoupled} and \eqref{eq:reduced_relax_ti_uncoupled}, the vCANNs is given by
    \begin{align}
        \Psi &= \Psi_1(\tilde{I}_1, \tilde{J}_1),  &G_1 &= G_1 \left( t; \tilde{I}_1, \tilde{J}_1\right).
    \end{align}
    \begin{table}[H]
        \centering
        \renewcommand{\arraystretch}{1.2}
        \begin{tabular}{ll}
            \toprule
            Hyperparameter & Value \\
            \midrule
            \emph{General} & \\
            \tabindent Learning rate &  $0.001$\\
            \tabindent Sparsity penalty parameter $\Lambda$ &  $0.0002$ \\
            \midrule
            \emph{Instantaneous elastic stress (CANN)} & \\
            \tabindent Convex & Yes\\
            \tabindent Number of neurons per hidden layer &  $\{ 32, 32, 16 \}$ \\
            \midrule
            \emph{Reduced relaxation functions} & \\
            \tabindent Maximal number of Maxwell elements $N_1^{max}$ & 10\\
            \tabindent Time normalization $[T_{min}, T_{max}]$ & $[10^{-2}, 10^{3}]$ s\\
            \tabindent Number of neurons per hidden layer of $\mathcal{N}_{\tau_{1\alpha}}$ & $\{ 32, 32, 16 \}$ \\
            \tabindent Number of neurons per hidden layer of $\mathcal{N}_{g_{1\alpha}}$ & $\{ 32, 32, 16 \}$ \\
            \bottomrule
        \end{tabular}
        \caption{Hyperparameters of the vCANN from Sec. \ref{sec:res_abdom}}
        \label{tab:hyper_abdom}
    \end{table}

    \subsection{Viscoelastic modeling of VHB 4910, Sec. \ref{sec:res_vhb4910}}\label{sec:hyper_vhb4910}
    The material is isotropic, exhibits strain-dependent but no strain rate-dependent viscous effects, and has no notable features ($\tns{f}=\tns{0}$). Thus, according to Eqs. \eqref{eq:SEF_ti_uncoupled} and \eqref{eq:reduced_relax_ti_uncoupled}, the vCANNs is given by
    \begin{align}
        \Psi &= \Psi_1(\tilde{I}_1, \tilde{J}_1),  &G_1 &= G_1 \left( t; \tilde{I}_1, \tilde{J}_1\right).
    \end{align}
    \begin{table}[H]
        \centering
        \renewcommand{\arraystretch}{1.2}
        \begin{tabular}{ll}
            \toprule
            Hyperparameter & Value \\
            \midrule
            \emph{General} & \\
            \tabindent Learning rate &  $0.001$\\
            \tabindent Sparsity penalty parameter $\Lambda$ &  $1.0$ \\
            \midrule
            \emph{Instantaneous elastic stress (CANN)} & \\
            \tabindent Convex & Yes\\
            \tabindent Number of neurons per hidden layer &  $\{8,8,6\}$ \\
            \midrule
            \emph{Reduced relaxation functions} & \\
            \tabindent Maximal number of Maxwell elements $N_1^{max}$ & 10\\
            \tabindent Time normalization $[T_{min}, T_{max}]$ & $[10^{-2}, 10^{3}]$ s\\
            \tabindent Number of neurons per hidden layer of $\mathcal{N}_{\tau_{1\alpha}}$ & $\{ 16, 16, 8 \}$ \\
            \tabindent Number of neurons per hidden layer of $\mathcal{N}_{g_{1\alpha}}$ & $\{ 16, 16, 8\}$ \\
            \bottomrule
        \end{tabular}
        \caption{Hyperparameters of the vCANN from Sec. \ref{sec:res_vhb4910}}
        \label{tab:hyper_vhb4910}
    \end{table}

    \subsection{Blast load analysis of Polyvinyl Butyral, Sec. \ref{sec:res_PVB}}\label{sec:hyper_PVB}
    The material is isotropic, exhibits strain-dependent and strain rate-dependent viscous effects, and has no notable features ($\tns{f}=\tns{0}$). Thus, according to Eqs. \eqref{eq:SEF_ti_uncoupled} and \eqref{eq:reduced_relax_ti_uncoupled}, the vCANNs is given by
    \begin{align}
        \Psi &= \Psi_1(\tilde{I}_1, \tilde{J}_1),  &G_1 &= G_1 \left( t; \tilde{I}_1, \tilde{J}_1, \tilde{\dot{I}}_1, \tilde{\dot{J}}_1, \mathrm{III}_{\dot{\tns{C}}}\right).
    \end{align}
    \begin{table}[H]
        \centering
        \renewcommand{\arraystretch}{1.2}
        \begin{tabular}{ll}
            \toprule
            Hyperparameter & Value \\
            \midrule
            \emph{General} & \\
            \tabindent Learning rate &  $0.0014$\\
            \tabindent Sparsity penalty parameter $\Lambda$ &  $0.025$ \\
            \midrule
            \emph{Instantaneous elastic stress (CANN)} & \\
            \tabindent Convex & Yes\\
            \tabindent Number of neurons per hidden layer &  $\{16,16,16\}$ \\
            \midrule
            \emph{Reduced relaxation functions} & \\
            \tabindent Maximal number of Maxwell elements $N_1^{max}$ & 10\\
            \tabindent Time normalization $[T_{min}, T_{max}]$ & $[10^{-2}, 10^{3}]$ s\\
            \tabindent Number of neurons per hidden layer of $\mathcal{N}_{\tau_{1\alpha}}$ & $\{ 24, 24, 24 \}$ \\
            \tabindent Number of neurons per hidden layer of $\mathcal{N}_{g_{1\alpha}}$ & $\{ 24, 24, 24 \}$ \\
            \bottomrule
        \end{tabular}
        \caption{Hyperparameters of the vCANN from Sec. \ref{sec:res_PVB}}
        \label{tab:hyper_PVB}
    \end{table}

    \subsection{Thermo-viscoelastic modeling of VHB 4905 data, Sec. \ref{sec:res_VHB4905}}\label{sec:hyper_vhb4905}
    The material is isotropic, exhibits strain-dependent but no strain rate-dependent viscous effects, and its mechanical behavior is significantly temperature-dependent ($\tns{f}=[\Theta]^{\mathrm{T}}$). Thus, according to Eqs. \eqref{eq:SEF_ti_uncoupled} and \eqref{eq:reduced_relax_ti_uncoupled}, the vCANNs is given by
    \begin{align}
        \Psi &= \Psi_1(\tilde{I}_1, \tilde{J}_1, \Theta),  &G_1 &= G_1 \left( t; \tilde{I}_1, \tilde{J}_1, \Theta \right).
    \end{align}
    \begin{table}[H]
        \centering
        \renewcommand{\arraystretch}{1.2}
        \begin{tabular}{ll}
            \toprule
            Hyperparameter & Value \\
            \midrule
            \emph{General} & \\
            \tabindent Learning rate &  $0.0005$\\
            \tabindent Sparsity penalty parameter $\Lambda$ &  $0.0001$ \\
            \midrule
            \emph{Instantaneous elastic stress (CANN)} & \\
            \tabindent Convex & Yes\\
            \tabindent Number of neurons per hidden layer &  $\{32,32,16\}$ \\
            \midrule
            \emph{Reduced relaxation functions} & \\
            \tabindent Maximal number of Maxwell elements $N_1^{max}$ & 10\\
            \tabindent Time normalization $[T_{min}, T_{max}]$ & $[10^{-2}, 10^{3}]$ s\\
            \tabindent Number of neurons per hidden layer of $\mathcal{N}_{\tau_{1\alpha}}$ & $\{ 32,32,16 \}$ \\
            \tabindent Number of neurons per hidden layer of $\mathcal{N}_{g_{1\alpha}}$ & $\{ 32,32,16 \}$ \\
            \bottomrule
        \end{tabular}
        \caption{Hyperparameters of the vCANN from Sec. \ref{sec:res_VHB4905}}
        \label{tab:hyper_vhb4905}
    \end{table}
    
	\section{VHB 4910 Data} \label{sec:VHB_data}
	\begin{figure}[H]
		\centering
		\includegraphics[width=0.7\linewidth]{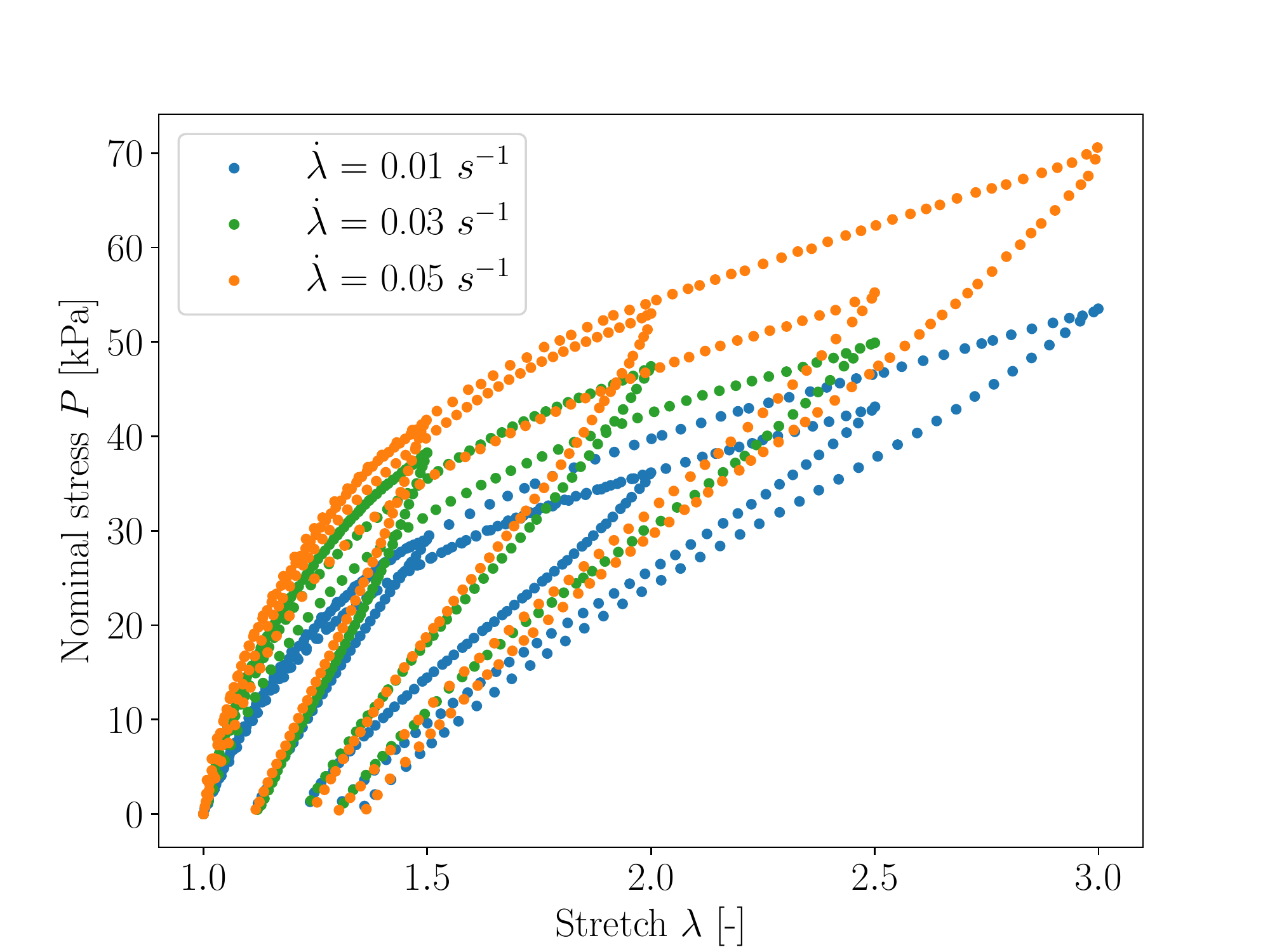}
		\caption{Uniaxial loading-unloading data of VHB 4910 taken from \cite{Hossain2012}. Note that for a given stretch rate $\dot{\lambda}$, the loading curves should coincide for all maximal stretches. However, for example, the loading curve with$\lambda_{max}=2.5$ and stretch rate $\dot{\lambda}=0.05$ $s^{-1}$ does not coincide with the other loading curves at $\dot{\lambda}=0.05$ $s^{-1}$. This suggests measurement errors or variations of the material samples not uncommon in mechanical testing. Such errors and variations in the data necessarily limit the extent to which a model can capture all the data.}
		\label{fig:vhb_data}
	\end{figure}

	\bibliography{library.bib}
	\bibliographystyle{unsrt}

\end{document}